\documentclass[prd,showpacs,amsmath,amssymb,nofootinbib,superscriptaddress]{revtex4}
\usepackage{array,mathtools,amssymb,booktabs,multirow,diagbox,hhline,natbib}
\pdfoutput=1

\usepackage{graphicx}
\usepackage{dcolumn}
\usepackage{bm}
\usepackage{bbold}
\usepackage{subfig}
\usepackage{graphicx}
\usepackage{amsmath}
\allowdisplaybreaks

\newcommand{\Od}{{\cal O}}
\newcommand{\tr}{\mbox{Tr}}
\newcommand{\diag}{\mbox{diag}}
\newcommand{\mean}[1]{\left\langle{#1}\right\rangle}
\newcommand{\condtwo}{\langle \bar q q \rangle}

\newcommand{\conds}{\langle \bar s s \rangle}
\newcommand{\condl}{\mean{\bar q q}_l}

\newcommand{\cth}{c_\theta}
\newcommand{\sth}{s_\theta}

\providecommand{\ec}{\;,}
\providecommand{\ep}{\;.}
\providecommand{\nt}{\notag}
\newcommand{\diff}{\text{d}}
\newcommand{\ID}{{\mathbb{1}}}

\begin{document}

\title{The QCD topological charge and its thermal dependence: the role of the $\eta'$}
\author{A. G\'omez Nicola}
\email{gomez@ucm.es}
\affiliation{Departamento de F\'{\i}sica
Te\'orica and IPARCOS. Univ. Complutense. 28040 Madrid. Spain}
\author{J. Ruiz de Elvira}
\email{elvira@itp.unibe.ch}
\affiliation{Albert Einstein Center for Fundamental Physics, Institute for Theoretical Physics,
University of Bern, Sidlerstrasse 5, CH--3012 Bern, Switzerland}
\author{A. Vioque-Rodr\'iguez}
\email{avioque@ucm.es}
\affiliation{Departamento de F\'{\i}sica
Te\'orica and IPARCOS. Univ. Complutense. 28040 Madrid. Spain}

\begin{abstract}
We analyze the contribution of the $\eta'(958)$ meson in the first two non-trivial moments of the QCD topological charge distribution, namely, the topological susceptibility and the fourth-order cumulant of the vacuum energy density.
We perform our study within U(3) Chiral Perturbation Theory up to next-to-next-to-leading order in the combined chiral and large-$N_c$ expansion. 
We also describe the temperature dependence of these two quantities and compare them with previous analyses in the literature. 
In particular, we discuss the validity of the thermal scaling of the topological susceptibility with the quark condensate, which is intimately connected with a Ward Identity relating both quantities.  
We also consider isospin breaking corrections from the vacuum misalignment at leading order in the U(3) framework. 
\end{abstract}

\maketitle

\section{Introduction and Motivation}
\label{sec:intro}

The rich topological structure of the QCD  vacuum is encoded in the $\theta$-angle dependence of the vacuum energy density 
\begin{equation}\label{evacdef}
  e_\text{vac}=-\frac{1}{\beta V} \ln Z(\theta)\ec\quad\text{with}\quad Z_\text{QCD}(\theta)=\int{[\diff G][\diff \bar\psi][\diff\psi]\,e^{S_\text{QCD}(\theta)}}
\end{equation}
the QCD generating functional in a $\theta$-vacuum, $\beta=1/T$ is the inverse of the temperature, $V$ the volume of the system and the QCD action is written as
\begin{equation}\label{Stheta}
S_\text{QCD}(\theta)=\int\diff^4 x \left[{\cal L}_\text{QCD}-\theta(x)\omega(x)\right]
\end{equation}
with $\cal{L}_\text{QCD}$ the QCD Lagrangian at $\theta=0$ and 
\begin{align}
  \omega(x)&=\frac{g^2}{32\pi^2}\mbox{Tr}_c G_{\mu\nu}\tilde G^{\mu\nu}\ec
  \label{ltheta}
\end{align}
the winding number topological charge density, responsible for the $U_A(1)$ anomaly.

The expansion of the vacuum energy density around $\theta=0$ can be expressed as 
\begin{equation}\epsilon_{vac} (\theta)=\sum_{n=1}^\infty \frac{c_{2n}}{(2n)!}\theta^{2n}\label{evac} \end{equation}
 with $c_2=\chi_{top}$ the topological susceptibility and $c_4$ the fourth-order cumulant, which in Euclidean space-time read
\begin{align}
\chi_{top}&=\int_T \diff x  \left\langle \mathcal{T} \omega(x) \omega(0) \right\rangle\ec\nt\\
c_4&=-\int_T \diff x\, \diff y\, \diff z\left[ \left\langle \mathcal{T} \omega(x) \omega(y)\omega(z) \omega(0)\right\rangle -3\left\langle \mathcal{T}\omega(x) \omega(0)\right\rangle^2\right]\ec
\label{chitopdef}
\end{align} 
with $\displaystyle \int_T \diff x=\int_0^\beta d\tau\int d^3{\vec{x}}$. 

The topological susceptibility is meant to be  connected with the $\eta'$ mass through the $U_A(1)$ anomaly. More specifically, in the quenched approximations where quarks loops are absent, 
$\chi_{top}^{quenched}$ is related to the mass of the $\eta'$ (or rather its singlet part $\eta_0$) for $N_f$ massless quarks as~\cite{Witten:1979vv,Veneziano:1979ec}
\begin{equation}
\chi_{top}^{\text{quenched}}=\frac{1}{2N_f}F^2 M_0^2=\frac{F^2\left[M^{m_q=0}_{\eta_0}\right]^2}{2N_f},
\label{chitopquenched}
\end{equation}
where $M_0$ denotes the anomalous contribution to the $\eta'$  mass and $F$ is the pion decay constant in the chiral $m_u=m_d=m_s=0$ limit. 
The quenched approximation is formally valid in the $N_c\rightarrow\infty$ limit, where meson-loop contributions are suppressed and the mass of the $\eta_0$ becomes of the same order than the other members of the Nambu-Goldstone boson (NGB) meson octet, since the $U(1)_A$ anomaly scales with $1/N_c$. 

However, even when the fluctuation of the winding number is directly linked to the $U(1)_A$ anomaly, meson-loop corrections are indeed very relevant and one of the reasons why the QCD topological charge can be analyzed using low-energy effective field theories. 
This  can be seen by looking at the leading-order (LO) low-energy chiral prediction for the topological susceptibility, 
a well-known result in Chiral Perturbation Theory (ChPT) that for three light flavors $N_f=2+1$ reads~\cite{DiVecchia:1980yfw,Gasser:1984gg,Leutwyler:1992yt}
\begin{equation}
\chi_{top}^{SU(3),LO}=\Sigma \left[\frac{1}{m_u}+\frac{1}{m_d}+\frac{1}{m_s}\right]^{-1}\equiv \Sigma\bar m,\label{chitopsu3lo}
\end{equation}
where $\Sigma=B_0F^2=-\condtwo$  is the single-flavor quark condensate in the chiral limit and $B_0=M_{0\pi^\pm}^2/(m_u+m_d)$, with $M_{0\pi^\pm}$ the mass of the charged pions. 
The result in~\eqref{chitopsu3lo} can be easily extended to a larger number of flavors~\cite{DiVecchia:1980yfw,Leutwyler:1992yt}, while the two-flavor result can be recovered by taking the $m_s>>m_{u,d}$ limit.  

One of the main consequences of~\eqref{chitopsu3lo} is that the topological susceptibility is linearly proportional to the quark mass and hence, unlike the pure gluonic quenched result~\eqref{chitopquenched}, it vanishes in the chiral limit. Thus, meson loop corrections generate terms that cancel the contribution in~\eqref{chitopquenched}. This fact is actually crucial to understand why the QCD topological charge can be reliable described just by including the lightest degrees of freedom.

In addition,~\eqref{chitopsu3lo} also shows that at LO in the chiral expansion the topological susceptibility is proportional to the quark condensate. 
This property has been used in several lattice analyses to extract the value of $\condtwo$ in the chiral limit from $\chi_{top}$, both for $N_f=2$~\cite{Aoki:2007pw,Bernardoni:2010nf} and $N_f=2+1$~\cite{Allton:2007hx,Chiu:2008jq}. 
In these analyses, $\chi_{top}$ is determined at different quark-mass values and then fitted using ChPT predictions to extract the quark condensate. 
This relation has been emphasized in~\cite{Bernard:2012fw} to study the strange mass paramagnetic suppression of the three-flavor condensate. 
Recent values of the quark condensate obtained with this method can be found in~\cite{Aoki:2016frl}, while direct lattice measurements of $\chi_{top}$ are provided in~\cite{Bonati:2015vqz,Petreczky:2016vrs,Borsanyi:2016ksw} for $N_f=2+1$, in~\cite{Dimopoulos:2018xkm} for $N_f=2$ and in \cite{Burger:2018fvb} within the framework of two fermion families.  
Therefore, obtaining explicit analytic expressions for $\chi_{top}$ and studying its quark mass dependence within the effective Lagrangian framework under different approximations is of the utmost importance from the point of view of lattice analyses. 

Nevertheless, higher order corrections change the simple linear dependence of $\chi_{top}$ with the quark condensate given in~\eqref{chitopsu3lo}. In this sense, an important result that hints towards more complicated dependencies is the existence of the following family of Ward Identities (WI) connecting quark condensates with the topological and pseudoscalar susceptibilities in the isospin limit~\cite{Hansen:1990yg,Nicola:2016jlj,Azcoiti:2016zbi,Nicola:2018vug}
\footnote{We acknowledge a misprint in eq~(31) of~\cite{Nicola:2018vug} where $\condl$ should read $\conds$.}:
\begin{align}
\chi_{top}&=-\frac{1}{4}\left[m_q\condl+m_q^2 \chi_P^{ll}\right] \label{wi1}\\
&=-\left[m_s\conds+m_s^2\chi_P^{ss}\right]\label{wi2}\ec
\end{align}
 where $\condl=\langle \bar u u + \bar d d\rangle$, $m_q=m_u=m_d$ and $\chi_P^{ll}$, $\chi_P^{ss}$ are the pseudoscalar susceptibilities corresponding to the quark bilinear operators $\eta_l=i\left(\bar u \gamma_5 u+\bar d \gamma_5 d\right)$ and $\eta_s=i\,\bar s s$, respectively. 

 The identities~\eqref{wi1} and~\eqref{wi2} have been verified in U(3) ChPT up to next-to-next-leading order (NNLO) in~\cite{Nicola:2016jlj}. Furthermore, the identity~\eqref{wi1} has been used in lattice works to determine $\chi_{top}$ indirectly at finite temperature~\cite{Buchoff:2013nra,Bhattacharya:2014ara} and in~\cite{GomezNicola:2017bhm} to justify that $U(1)_A$ restoration approaches the $O(4)$ phase transition when the chiral symmetry is exactly restored. 

Now, note that for $N_f=2$ in the isospin limit, the first term in the r.h.s. of~\eqref{wi1} corresponds to the LO ChPT expression~\eqref{chitopsu3lo} when $m_s>>m_q$. Therefore, the term proportional to $\chi_P^{ll}$ necessarily includes  higher order corrections in the chiral series and/or terms suppressed as $m_q/m_s$ for $N_f=2+1$ flavors.  An immediate conclusion is that $\chi_{top}$ is not necessarily proportional to the light quark condensate to all orders. In fact, as we will see here in detail, the dependence with temperature of the two terms in the r.h.s. of~\eqref{wi1} is completely different. 
Namely, the condensate term drops with $T$ signaling chiral restoration while the $\chi_P^{ll}$ term shows a much smoother behavior.  

Higher order corrections to the topological susceptibility within the chiral Lagrangian framework have been obtained in~\cite{Mao:2009sy} to next-to-leading order (NLO) (one-loop) in  SU(N$_{\rm f}$) ChPT. 
This result has been incorporated in the lattice analysis of~\cite{Bernardoni:2010nf} to extract the quark condensate and in~\cite{diCortona:2015ldu,Gorghetto:2018ocs} to provide a numerical estimate of $\chi_{top}$ in terms of the low-energy constants (LECs) of the $\Od(p^4)$ effective Lagrangian. Higher order isospin-breaking corrections as well as an analysis of $\chi_{top}$ within the  so called resummed ChPT can be found in~\cite{Bernard:2012fw}, whereas in the recent work~\cite{Gorghetto:2018ocs} NNLO and electromagnetic corrections in SU(2) ChPT have been computed. 

Another important motivation for the study of the topological susceptibility is its relation with the Peccei-Quinn axion~\cite{Peccei:1977hh,Weinberg:1977ma} and hence with various cosmological and astrophysical implications. The axion mass is directly proportional to $\chi_{top}$, which allows for  numerical estimates based on chiral Lagrangians~\cite{Spalinski:1988az,diCortona:2015ldu,Gorghetto:2018ocs}. In addition, the fourth-order self-coupling of the axion field can be obtained from the fourth-order cumulant $c_4$ of the $\epsilon_{vac}(\theta)$ expansion in~\eqref{evac}, which has been thoroughly analyzed  within SU(N$_{\rm f}$) ChPT up to NLO~\cite{Mao:2009sy,Bernard:2012ci,Guo:2015oxa} and computed in the lattice~\cite{Chiu:2008jq,Vicari:2008jw,Panagopoulos:2011rb,Bonati:2016tvi}.  

In addition, the large-$N_c$ behavior of the topological susceptibility and the fourth-order cumulant have been recently analyzed in \cite{Vonk:2019kwv} and compared with the lattice results provided at different large-$N_c$ values in~\cite{Vicari:2008jw}.

The thermal dependence of the vacuum energy density  $\epsilon(\theta)$ is important for several reasons:  as stated above $\chi_{top}(T)$ plays an important role in the relation between chiral and $U(1)_A$ restoration. In addition, the connection of $\chi_{top}$ with the light- and strange-quark condensate as a function of temperature is relevant for lattice analyses and for the understanding of the temperature dependence of the different contributions in the WI~\eqref{wi1}-\eqref{wi2}.  Thermal corrections to the axion potential and its mass are also of importance in the cosmological and astrophysical context~\cite{diCortona:2015ldu}.  
A ChPT analysis for $\chi_{top}(T)$ for $N_f=2$ has been performed in~\cite{diCortona:2015ldu}. In that paper, the fact that for $N_f=2$ the one-loop corrections to $\chi_{top}$ can be  encoded in the physical pion mass and  decay constant is used to establish the scaling $\chi_{top}(T)/\chi_{top}(0)=\condl (T)/\condl (0)$, valid  at that order. Actually, that scaling law is nothing but the first term in the r.h.s. of~\eqref{wi1}, which opens the question of how relevant are the additional corrections provided by the second term in that identity. As we have already mentioned, we will discuss that particular aspect in detail in the present  work.

As for lattice results at finite temperature, direct measurements of the topological susceptibility have typically large errors. Results can be found  e.g. in~\cite{Bazavov:2012qja,Borsanyi:2016ksw,Bonati:2015vqz,Petreczky:2016vrs}. As commented above, indirect measurements can be obtained precisely through~\eqref{wi1}. Higher order cumulants at finite temperature in the lattice have been analyzed in~\cite{Bonati:2013tt,Bonati:2015vqz}.

With the above motivation in mind, we will carry out here a ChPT-based analysis of the topological susceptibility and the fourth-order cumulant, concentrating in particular in the following aspects: 
\begin{itemize}
\item We will provide a NNLO calculation of the topological susceptibility and the fourth-order cumulant within the formalism of U(3) ChPT, in which the singlet $\eta_0$ is incorporated  as a ninth pseudo-Goldstone boson within the large-$N_c$ framework ~\cite{Witten:1979vv,HerreraSiklody:1996pm,Kaiser:2000gs}. This approach allows us to study $\eta'$ meson effects and to assess its contribution. For instance, the inclusion of $\eta'$ will help to understand the relevance of meson-loop corrections to the quenched result~\eqref{chitopquenched} and their role on the vanishing of $\chi_{top}$ in the chiral limit, as already noticed in~\cite{Leutwyler:1992yt}.
 Our calculation will also allow us to estimate numerically the effect of the additional U(3) corrections in terms of the LECs involved. As mentioned above, many lattice analyses of the quark condensate rely on the chiral expansion of $\chi_{top}$. In that respect, studying the influence of an additional heavier degree of freedom is important. The $\eta'$ case is especially significant due to its direct connection with the axial anomaly and its relevant role in the Ward identities described above. In addition, our $U(3)$ analysis will provide a natural way to establish contact with recent large-$N_c$ analysis of the topological susceptibility and the fourth-order cumulant. Our study will also have the advantage of including explicitly the dependence of those quantities on the $\eta-\eta'$ mixing angle. 

\item  We will also calculate the leading isospin-breaking $m_u\neq m_d$ corrections within the U(3) formalism, thus extending previous SU(3) works. The importance of isospin breaking corrections to the topological susceptibility and the fourth-order cumulant stems from the vacuum misalignment induced by the combined $m_u\neq m_d$ and $\theta\neq 0$ effects~\cite{DiVecchia:1980yfw,Gasser:1984gg,Mao:2009sy,Guo:2015oxa,Gorghetto:2018ocs}. It implies corrections proportional to $(m_u-m_d)/(m_u+m_d)$, hence much larger than the typical isospin-breaking correction in other quantities, such as quark condensates or the $\pi^0\eta$ mixing~\cite{Nicola:2010xt}, which are proportional to $(m_u-m_d)/m_s$.
Recent estimates within the SU(2) formalism show that these isospin contributions give rise to around a 4\% correction to $\chi_{top}^{1/4}$. 
As a natural extension to those analysis, we will include isospin breaking in the LO U(3) correction to $\epsilon_{vac}(\theta)$ to estimate its numerical effect.

\item We will extend our analysis to finite temperature. In this way, we will account for corrections both within the SU(3) and U(3) formalisms for the different contributions in the Ward identities~\eqref{wi1} and~\eqref{wi2}. This will also allow us to test the robustness of the $N_f=2$ scaling  performed in~\cite{diCortona:2015ldu} when corrections from $m_q /m_s$ and $\eta,\eta'$ loops are properly incorporated. As commented before, the evolution of  quark condensates and susceptibilities towards chiral restoration makes it interesting to clarify their relation with the topological susceptibility as the temperature grows, within the context of chiral and $(1)_A$ restoration.  In addition, we recall that the finite temperature dependence of the $\eta'$ mass has been analyzed within the $U(3)$ ChPT formalism in~\cite{Escribano:2000ju,Gu:2018swy},  in fermion models~\cite{Ishii:2016dln} and in the lattice~\cite{Kotov:2019dby}, confirming the $U(1)_A$ restoring behavior, as in the recent analysis of the $\eta-\eta'$ mixing angle \cite{Nicola:2018vug}.

\end{itemize}

For that purpose, the paper is organized as follows. In section~\ref{sec:u3} we will discuss the U(3) ChPT calculation of the topological susceptibility, providing explicit analytic expressions up to NNLO. 
We will also analyze its various limits of interest. In section~\ref{sec:fourth} we will extend the calculation to the fourth order cumulant. In Section~\ref{sec:num} we will provide numerical estimates and compare to previous approaches. The isospin breaking corrections to the U(3) susceptibility and the fourth-order cumulant will be calculated in section~\ref{sec:ib}. In section~\ref{sec:temp} we will study in detail their finite temperature dependence and  their connection with chiral and $U(1)_A$ restoration. Some of the explicit analytic U(3) expressions will be collected in Appendix~\ref{app:results}.

\section{The  topological susceptibility to NNLO in U(3) Chiral perturbation Theory}
\label{sec:u3}


Within U(3) ChPT one follows a similar approach as in standard ChPT~\cite{Gasser:1984gg}. The chiral power counting in terms of momenta and quark masses is used to construct the most general effective Lagrangian for SU(N$_{\rm f}$) pseudo-Goldstone bosons up to a given order, which ensures renormalizability order by order in the expansion. 
Nevertheless, in the U(3) formalism the singlet $\eta_0$ is also included as the ninth pseudo-Goldstone boson. Given the large $\eta'$ mass value, this can be done consistently only in the large-$N_c$ framework, 
since the winding number charge density $\omega(\theta)$ defined in~\eqref{ltheta}, responsible for the anomalous contribution of the $\eta'$ mass, is suppressed within the large-$N_c$ counting~\cite{Witten:1979vv,DiVecchia:1980yfw,Rosenzweig:1979ay,HerreraSiklody:1996pm,Kaiser:2000gs}. 

Thus, the expansion is performed in terms of a small parameter $\delta$ such that $\{M^2, p^2, T^2, m_q,m_s,1/N_c\}=\Od(\delta)$, where $M$ and $p$ denote typical meson masses and momenta. 
In this counting, the tree-level pion decay constant $F^2=\Od(N_c)=\Od(1/\delta)$, which hence suppresses loop diagrams. 
The counting of the different LECs according to their $\Od(N_c)$ trace structure is given in detail in~\cite{HerreraSiklody:1996pm,Guo:2012ym,Guo:2015xva}. 

Following the same steps as in~\cite{GomezNicola:2010tb}, the topological susceptibility can be calculated by taking functional derivatives with respect to the vacuum angle $\theta$. 
Thus, taking into account the $\theta$-vacuum coupling in the QCD action defined in~\eqref{Stheta} and \eqref{ltheta}, 
one can derive expectation values or thermal correlators involving the winding number density. In this way, the topological susceptibility reads
 \begin{align}
 \chi_{top}&=\int_T \diff x \left.\frac{\delta}{\delta \theta (x)}\frac{\delta}{\delta \theta (0)}\log Z(\theta)\right\vert_{\theta=0}\nt\\
&=\int_T \diff x\, \left\{ \left\langle\frac{\delta{\cal L}_{\rm eff} (x)}{\delta \theta (x)}\frac{\delta {\cal L}_{\rm eff} (0)}{\delta \theta (0)}\right\rangle_{\theta=0} +\left\langle \frac{\delta}{\delta \theta (x)}  \frac{\delta}{\delta \theta (0)}  {\cal L}_{\rm eff} (x)  \right\rangle_{\theta=0} \delta^4(x)\right\}\ec
 \label{functexpchi}
 \end{align}
 where $\langle\cdot\rangle$ denotes Euclidean vacuum expectation values for $T=0$ or thermal correlators for $T\neq 0$ and where we have used that $\langle \omega(0)\rangle=0$. 
 
 In the effective Lagrangian ${\cal L}_{\rm eff}$, $\theta(x)$ appears through the operator 
\begin{align}\label{Xfield}
X(x)=\log\left[\det U(x)\right] + i\theta(x),
\end{align}
with $U=\exp(i\Phi/F)=\exp(i\lambda^a\phi_a/F)$ the NGB matrix field including the singlet contribution (i.e., $\det U\neq 1$), $\phi_i$ the Goldstone fields,  $\lambda^{a=1,\dots 8}$ the Gell-Mann matrix and $\lambda^0=\sqrt{2/3} \ \ID$~\cite{HerreraSiklody:1996pm},
\begin{equation}\label{phi1}
\Phi \,=\, \left( \begin{array}{ccc}
\pi^0+\frac{1}{\sqrt{3}}\eta_8+\sqrt{\frac{2}{3}} \eta_0 & \sqrt{2}\pi^+ & \sqrt{2}K^+ \\ \sqrt{2}\pi^- &
-\pi^0+\frac{1}{\sqrt{3}}\eta_8+\sqrt{\frac{2}{3}} \eta_0   & \sqrt{2}K^0 \\  \sqrt{2}K^- & \sqrt{2}\bar{K}^0 &
\frac{-2}{\sqrt{3}}\eta_8+\sqrt{\frac{2}{3}} \eta_0 
\end{array} \right)\,.
\end{equation}
Nevertheless, due to the $\eta-\eta'$ mixing the flavor eigenstates $\eta_8$ and $\eta_0$ are not mass eigenstates even at LO in the chiral Lagrangian. 
Thus, we use the angle $\hat\theta$ to describe their mixing at LO
\begin{equation}
\eta_8= \cth\,\eta +\sth\,\eta'\quad\eta_0= -\sth\,\eta +\cth\,\eta',
\end{equation}
where $\cth=\cos\hat\theta$ and $\sth=\sin\hat\theta$. 

The combination~\eqref{Xfield} ensures that ${\cal L}_{\rm eff}$ will be invariant under local $U(N_f)_L\times U(N_f)_R$ transformations. 
The effective Lagrangians containing the $X$ field entering in $\chi_{top}$ and the fourth-order cumulant up to the order needed for our purposes here read
 \begin{align}
 \Od(\delta^0):& \qquad {\cal L}_{\rm eff}^{(0)}=\frac{F^2}{12}\,M_0^2\, X^2 \label{efflag0}\ec\\
 \Od(\delta):& \qquad {\cal L}_{\rm eff}^{(1)}=\frac{-F^2}{12}\Lambda_2 X\,  \tr\left(U^\dagger\chi-\chi^\dagger U\right)\label{efflag1}\ec\\
 \Od(\delta^2):& \qquad {\cal L}_{\rm eff}^{(2)}= \frac{F^2}{4}\left[v_4^{(0)}X^4+v_2^{(2)} X^2\, \tr\left(U^\dagger\chi+\chi^\dagger U\right)\right] +L_{25} X\, \tr\left( U^\dagger \chi U^\dagger \chi-\chi^\dagger U \chi^\dagger U\right)\ec
 \label{efflag2}
 \end{align}
where  $M_0^2$ is  the contribution to the tree-level mass of the singlet $\eta_0$ in the chiral limit, i.e., its anomalous contribution as given in~\eqref{chitopquenched},  and $\chi=2B_0{\cal M}$ with ${\cal M}=\diag(m_u,m_d,m_s)$ the quark mass matrix. The constants $\Lambda_2=\Od(\delta)$, $v_2^{(2)}=\Od(\delta^2)$, $v_4^{(0)}=\Od(\delta^3)$  and $L_{25}=\Od(\delta^0)$ are LECs associated to the $\eta-\eta'$ mixing. As in the standard SU(3) formalism, the LECs  are renormalized to absorb divergences coming from the loops. 

The first non-vanishing contribution to the topological susceptibility is $\Od(\delta^0)$. On the one hand, the second term in the r.h.s. of~\eqref{functexpchi} gives rise at this order to a constant term, 
which is nothing but the contribution~\eqref{chitopquenched} with $\left[M^{m_q=0}_{\eta_0}\right]^2=M_0^2$ and $N_f=3$. 
On the other hand, the first term in the r.h.s of~\eqref{functexpchi}  gives rise to terms of the type  $M_0^4\int \diff x\langle \eta^{(')}(x)\eta^{(')}(0)\rangle\sim M_0^4/M_{\eta^{(')}}^2$, 
i.e., tree level two-point functions at $p=0$,  where $\eta^{(')}$ denotes generically $\eta$ or $\eta'$ fields.  Gathering the two types of contribution yields in the isospin limit
\begin{equation}
\chi_{top}^{U(3),\, LO,\, IL}=\frac{F^2M_0^2}{6}\left[1-\frac{M_0^2}{M_{0\eta}^2}\sth^2-\frac{M_0^2}{M_{0\eta'}^2} \cth^2  \right]\ec
\label{chitopu3lo}
\end{equation}
where $M_{0\eta}$ and $M_{0\eta'}$ are respectively the $\eta$ and $\eta'$ masses at tree level. 
They depend on the quark (or meson) masses and on $M_0$. Their explicit expressions in the isospin limit $m_u=m_d=m_q$ can be found  e.g. in~\cite{Guo:2015xva}. 
Thus, in terms of quark masses,~\eqref{chitopu3lo} can be recast as:
\begin{equation}
\chi_{top}^{U(3),\, LO,\, IL}=\frac{F^2M_0^2 B_0 m_q m_s}{(2m_s+ m_q)M_0^2+6B_0 m_q m_s }=\frac{\Sigma M_0^2 {\bar m}_{IL}}{M_0^2+6B_0 {\bar m}_{IL}}=\Sigma {\hat m}_{IL}\ec
\label{chitopu3lob}
\end{equation}
where ${\bar m}_{IL}$ denotes the isospin limit value of  $\bar m$  in~\eqref{chitopsu3lo}, i.e.,  
 \begin{equation}
  {\bar m}_{IL}=\frac{m_q m_s}{2m_s+m_q}\ec\label{eq:barm}
\end{equation}
   and
 \begin{equation}
{\hat m}_{IL}=  \frac{M_0^2 {\bar m}_{IL}}{M_0^2+6B_0 {\bar m}_{IL}}\ep\label{eq:hatm}
\end{equation}

 Note that in this work we use the symbol $\hat m$, as in~\eqref{eq:hatm}, with a different meaning than the light-quark mass, which we denote here by  $m_q$.  The result in~\eqref{chitopu3lo} and~\eqref{chitopu3lob} is the extension to $N_f=2+1$ flavors of the  result given in~\cite{Leutwyler:1992yt} for $N_f$ degenerated flavors in the large-$N_c$ limit. 
Note that our U(3) LO result can be obtained from the SU(3) one by replacing $\bar{m}$ with $\hat{m}$, which also holds when including isospin breaking corrections at LO, as we will see in section~\ref{sec:ib}. 
In the $M_0\rightarrow\infty$ limit one recovers the SU(3) expression in~\eqref{chitopsu3lo} for $m_u=m_d=m_q$, since ${\hat m}_{IL}\rightarrow {\bar m}_{IL}$ in this limit.
Furthermore, $\chi_{top}^{U(3),LO}$ vanishes in the chiral limit, as it also does in the SU(3) case.  
Actually, in this limit the second and third terms in the r.h.s. of~\eqref{chitopu3lo} (coming from $\eta,\eta'$ propagators) cancel the first term, as one can check from the chiral limit behavior of $s_\theta,M_{0\eta},M_{0\eta'}$.
 
Our present U(3) calculation has also the advantage that one can formally recover the quenched result in~\eqref{chitopquenched} by taking the limit $M_0\ll \bar m$. 
This limit can also be achieved by re-expanding~\eqref{chitopu3lob} in the $1/N_c$ expansion. Taking into account the $1/N_c$ scaling of the different constants involved, namely $F^2=\Od(N_c)$, $M_0^2=\Od(1/N_c)$ and so on, the $1/N_c$ LO contribution to~\eqref{chitopu3lob} gives
 \begin{equation}
 \chi_{top}^{U(3),IL}=\frac{F^2 M_0^2}{6}   + \Od\left(\frac{1}{N_c}\right)=  \frac{F^2}{6}  \left(M_{0\eta'}^2+M_{0\eta}^2-2M_{0K}^2\right)  + \Od\left(\frac{1}{N_c}\right)\ec
 \label{largenclo}
 \end{equation}
 which coincides with the result in~\eqref{chitopquenched} for $N_f=3$.
 The last equality in~\eqref{largenclo} reproduces the result given in~\cite{Veneziano:1979ec}, where $M_{0i}$ denote tree-level meson masses. 
 
The NLO $\left(\Od(\delta)\right)$ and NNLO $\left(\Od(\delta^2)\right)$ results require including the higher-order effective Lagrangians~\eqref{efflag1}-\eqref{efflag2} in~\eqref{functexpchi}. They involve one-loop corrections to the $\eta$- and $\eta'$-meson propagators at zero momentum, which include the LECs $L_6$, $L_7$, $L_{25}$, $C_{19}$, $C_{31}$, $\Lambda_2$ and $v_2^{(2)}$~\cite{Guo:2015xva}. The renormalization of these LECs and the constant $B_0$ \cite{HerreraSiklody:1996pm,Kaiser:2000gs, Guo:2015xva} allows one to absorb all one-loop divergences, rendering the result finite and independent of the renormalization scale $\mu$. In addition, the calculation of $\chi_{top}$, $\chi_P^{ll}$, $\chi_P^{ss}$ and $\condl$ up to NNLO in the U(3) expansion allow one to verify the Ward Identities~\eqref{wi1}-\eqref{wi2}.  The explicit results for the topological susceptibility at NLO and NNLO in U(3) ChPT are given explicitly in Appendix~\ref{app:results} in the isospin limit. Note that the NLO and NNLO U(3) results do not correspond to simply perform the replacement $\bar m \rightarrow \hat{m}$ as it happened at LO.  

In order to compare with the SU(3) calculation in ChPT, as given for instance in~\cite{Mao:2009sy,diCortona:2015ldu,Bernard:2012fw}, we recall that the NLO order ChPT result is included distributed among the NLO and NNLO U(3) outcome. On the one hand, we have checked that the $M_0\rightarrow\infty$ limit of $\chi_{top}^{U(3),NLO}$ in~\eqref{chitopu3nlo} yields the contribution proportional to the renormalized LEC $L_8^r$ in~\cite{Mao:2009sy,diCortona:2015ldu,Bernard:2012fw} while  $\chi_{top}^{U(3),NNLO}$ in~\eqref{chitopu3nnlo} for $M_0\rightarrow\infty$ provides the rest of the SU(3) contributions, proportional to $L_6^r$, $L_7$ and $\log\left(M^2_{\pi,K,\eta}/\mu^2\right)$.  On the other hand, the surviving term proportional to $\log\left(M^2_{\eta'}/\mu^2\right)$ is absorbed in  $L_6^r$. 

Finally, we remark that our present U(3)  formalism allows us to study  systematically the large $N_c$ corrections in~\eqref{largenclo}, corresponding to the Witten-Veneziano result. Performing the $1/N_c$ expansion on the different orders in the U(3) ChPT expansion we obtain the $\Od(1/N_c)$ correction to~\eqref{largenclo}, namely
\begin{eqnarray}
\chi_{top}^{U(3),IL}&=&\frac{F^2 M_0^2}{6}  \left\{1+\frac{1}{3} M_0^2 \left(\frac{1}{M_{0\pi }^2-2 M_{0K}^2}-\frac{2}{M_{0\pi }^2}\right)+\frac{16\, M_0^2 L_8^r}{F^2}-2 \Lambda _2 + \frac{2\, \left(2 M_{0K}^2+M_{0\pi }^2\right)}{3 F^4 M_0^2} \left[24\, C_{19} F^2 M_0^4 \right.\right.\nonumber\\
&+&\left.\left. 16\, C_{31} F^2 M_0^4+9\, F^4 v_2^{(2)}+16\, F^2 \Lambda _2 M_0^2 L_8^r+24\, F^2 M_0^2 L_{25}^r-128\, M_0^4 (L_8^r)^2\right]\right\} + \Od\left(\frac{1}{N_c^2}\right)+ \Od\left(\delta^3\right)\ep
\label{largenc}
\end{eqnarray}


The above result is consistent with the large-$N_c$  scaling analysis of  the topological susceptibility provided in \cite{Vonk:2019kwv}. 

\section{Fourth-order cumulant to NNLO in U(3) ChPT}
\label{sec:fourth}

\begin{figure}
  \includegraphics[width=0.99\textwidth]{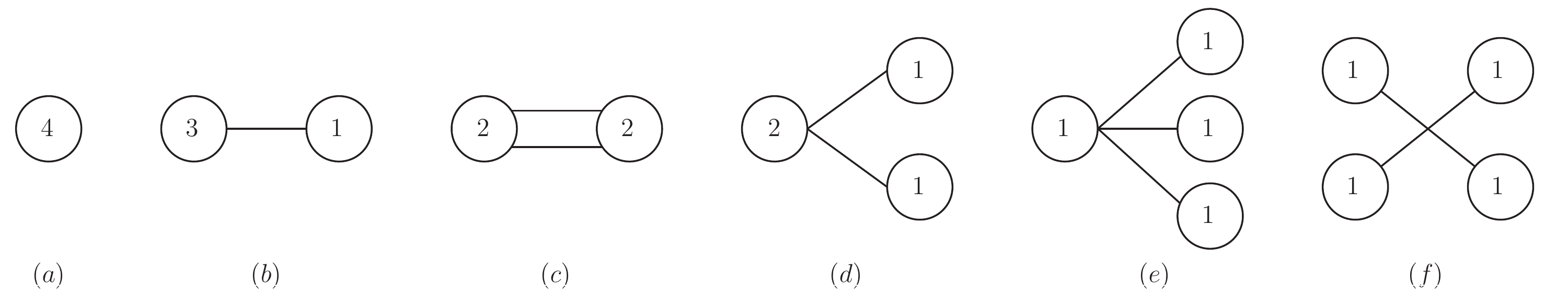}
  \caption{LO topologies for each of the five different connected contributions of the fourth-order cumulant in~\eqref{functexpc4}.
    Following the notation in~\cite{Bernard:2012ci}, $\theta$-induced vertices are denoted by its number of derivatives with respect to $\theta$ applied to the effective Lagrangian ${\cal L}_{\rm eff}$. Note that we only refer to the LO  for each different connected topology, irregardless of their corresponding counting in the $\delta$  expansion.
}
  \label{fig:c4topologies}
\end{figure}

The fourth-order cumulant is defined in~\eqref{chitopdef} and involves the difference between the four- and two-point function square of the winding number density. 
Similarly to $\chi_{top}$, it can be computed by taking functional derivatives with respect to the $\theta$-vacuum angle. 
Taking into account again the $\theta$-angle coupling in the QCD action as defined in~\eqref{Stheta} and \eqref{ltheta}, the fourth-order cumulant can be written as
 \begin{align}
 c_{4}=&-\int_T \diff x\,\diff y\,\diff z \left[\left.\frac{\delta}{\delta \theta (x)}\frac{\delta}{\delta \theta (y)}\frac{\delta}{\delta \theta (z)}\frac{\delta}{\delta \theta (0)}\log Z(\theta)\right\vert_{\theta=0}-3\left(\left.\frac{\delta}{\delta \theta (x)}\frac{\delta}{\delta \theta (0)}\log Z(\theta)\right\vert_{\theta=0}\right)^2\right]=\nt\\[1em]
=&-\Bigg[\left\langle\frac{\delta^4 {\cal L}_{\rm eff} (0)}{\delta \theta (0)^4}\right\rangle_{\theta=0}+\int_T \diff x\left\{4\left\langle\frac{\delta^3 {\cal L}_{\rm eff} (x)}{\delta \theta (x)^3}\frac{\delta{\cal L}_{\rm eff} (0)}{\delta \theta (0)}\right\rangle_{\theta=0}+3\left\langle\frac{\delta^2 {\cal L}_{\rm eff} (x)}{\delta \theta (x)^2}\frac{\delta^2{\cal L}_{\rm eff} (0)}{\delta \theta (0)^2}\right\rangle_{\theta=0}-3\left\langle\frac{\delta^2 {\cal L}_{\rm eff} (x)}{\theta (x)^2}\right\rangle_{\theta=0}^2\right\}\nt\\[0.8em]
&\quad+6\int_T \diff x\,\diff y \left\{\left\langle\frac{\delta^2{\cal L}_{\rm eff} (x)}{\delta \theta (x)^2}\frac{\delta{\cal L}_{\rm eff} (y)}{\delta \theta (y)}\frac{\delta{\cal L}_{\rm eff} (0)}{\delta \theta (0)}\right\rangle_{\theta=0}-\left\langle\frac{\delta^2 {\cal L}_{\rm eff} (x)}{\delta\theta (x)^2}\right\rangle_{\theta=0}\left\langle\frac{\delta{\cal L}_{\rm eff} (y)}{\delta\theta (y)}\frac{\delta{\cal L}_{\rm eff} (0)}{\delta\theta (0)}\right\rangle_{\theta=0}\right\}\nt\\[0.8em]
&\quad+\int_T \diff x\,\diff y\,\diff z\left\{\left\langle\frac{\delta{\cal L}_{\rm eff} (x)}{\delta \theta (x)}\frac{\delta{\cal L}_{\rm eff} (y)}{\delta \theta (y)}\frac{\delta{\cal L}_{\rm eff} (z)}{\delta \theta (z)}\frac{\delta{\cal L}_{\rm eff} (0)}{\delta \theta (0)}\right\rangle_{\theta=0}-3\left\langle\frac{\delta{\cal L}_{\rm eff} (x)}{\delta\theta (x)}\frac{\delta{\cal L}_{\rm eff} (0)}{\delta\theta (0)}\right\rangle_{\theta=0}^2\right\}\Bigg]\ec
 \label{functexpc4}
 \end{align}
where we have used once more that  $\langle \omega(0)\rangle=0$. The last terms in all rows are associated to the square of $\chi_{top}$ and hence they provide disconnected contributions, i.e., terms proportional to the Euclidean four-dimensional volume that should cancel out exactly with the disconnected contributions coming from the four-point function.
Thus, the calculation of the fourth-order cumulant involves five connected contributions associated to all possible combinations of a total even number of derivatives with respect to $\theta$  applied to the effective lagrangian.
The LO topologies for each of these five connected contributions are depicted in Fig.~\ref{fig:c4topologies}\footnote{Note that we only refer about the LO topology for each connected contribution in~\eqref{functexpc4} independently of their corresponding counting in the U(3) expansion.}.
Following the notation given in~\cite{Bernard:2012ci}, the $\theta$-induced vertices are classified in terms of their number of derivatives with respect to the $\theta$-vacuum angle.
Meson lines are always coupled to one $\theta$-induced source coming from a single, second or third derivative with respect to the vacuum angle~\footnote{The fourth derivative is just a contact term without any meson line up to NNLO in the $\delta$ expansion.}, which in the isospin limit involve $\eta$ and $\eta'$ meson propagators at zero momentum. The four different $\theta$-induced vertices contributing to the topologies in Fig.~\ref{fig:c4topologies} read
\begin{align}
\left.\frac{\delta {\cal L}_{\rm eff} (x)}{\delta \theta (x)}\right\vert_{\theta=0}=&-\frac{F}{\sqrt 6}\Bigg[M_0^2\left( \cth\, \eta'(x)-\sth\, \eta(x)\right)+\frac{2}{3}\Lambda_2B_0\Big((2 m_q+m_s)\left( \cth\, \eta'(x)-\sth\, \eta(x)\right)\nt\\
&\quad+\sqrt 2(m_q-m_s)\left( \cth\, \eta'(x)+\sth\, \eta(x)\right)\Big)+12B_0v_2^{(2)}(2m_q+m_s)\left( \cth\, \eta'(x)-\sth\, \eta(x)\right)\Bigg]\nt\\
&\quad+16\sqrt{\frac{2}{3}}\frac{B_0^2}{F}L_{25}\left[\left(2m_q^2+m_s^2\right)\left( \cth\, \eta'(x)-\sth\, \eta(x)\right)+\sqrt 2(m_q^2-m_s^2)\left( \cth\, \eta'(x)-\sth\, \eta(x)\right)\right]\nt\\
&\quad+\frac{1}{9\sqrt 3F}B_0\Lambda_2\Bigg[\frac{1}{3} \left((m_q - 4 m_s)\cth^3 -3 \sqrt2 (m_q + 2 m_s) \cth^2 \sth + 6 (m_q - m_s) \cth \sth^2 - \sqrt2 (2 m_q + m_s) \sth^3)\right)\eta(x)^3\nt\\
&\quad+\left(\sqrt2 (m_q + 2 m_s)\cth^3  - 3  m_q\cth^2\sth -3 \sqrt2  m_s \cth \sth^2 + 2 (m_q - m_s) \sth^3)\right)\eta(x)^2\eta'(x)\nt\\
&\quad+\left(2 (m_q - m_s)\cth^3  + 3  m_s\cth^2\sth -3 \sqrt2  m_q \cth \sth^2 - \sqrt2 (m_q + 2 m_s) \sth^3)\right)\eta(x)\eta'(x)^2 \nt\\
&\quad+\frac{1}{3} \left(\sqrt2 (2 m_q + m_s) \cth^3 +6 (m_q - m_s) \cth^2 \sth + 3 \sqrt2 (m_q + 2 m_s) \cth \sth^2 + (m_q - 4 m_s)\sth^3)\right)\eta'(x)^3\Bigg]\nt\\
&\quad+\frac{2}{F}\sqrt{\frac{2}{3}}B_0v_2^{(2)}\left( \cth\, \eta'(x)-\sth\, \eta(x)\right)\Bigg[\Big((m_q + 2 m_s)\cth^2- 2\sqrt2(m_q - m_s)\cth\sth+(2 m_q + m_s)\sth^2)\Big)\eta(x)^2\nt\\
&\quad +2(m_q - m_s)\left(\sqrt2 \cth^2-\cth\sth-\sqrt2\sth^2)\right)\eta(x)\eta'(x)+\Big((2m_q + m_s)\cth^2+2\sqrt2(m_q - m_s)\cth\sth\nt\\
&\quad+(m_q + 2m_s)\sth^2)\Big)\eta'(x)^2\Bigg]+\frac{6\sqrt 6}{F} v_4^{(4)}\left( \cth\, \eta'(x)-\sth\, \eta(x)\right)^3+\cdots+\Od\left(\delta^{9/2}\right)\label{eq:c4d1}\\
\left.\frac{\delta^2 {\cal L}_{\rm eff} (x)}{\delta \theta (x)^2}\right\vert_{\theta=0}=&-F^2\left[\frac{M_0^2}{6}+ (2m_q +m_s) B_0 v_2^{(2)}\right]+\frac{2}{3}v_2^{(2)}B_0\Bigg[\Big((m_q + 2 m_s)\cth^2- 2\sqrt2(m_q - m_s)\cth\sth+(2 m_q + m_s)\sth^2)\Big)\eta(x)^2\nt\\
&\quad +2(m_q - m_s)\left(\sqrt2 \cth^2-\cth\sth-\sqrt2\sth^2)\right)\eta(x)\eta'(x)+\Big((2m_q + m_s)\cth^2+2\sqrt2(m_q - m_s)\cth\sth\nt\\
&\quad+(m_q + 2m_s)\sth^2)\Big)\eta'(x)^2\Bigg]+18v_4^{(0)}\left(\cth \eta'(x)-\sth \eta(x)\right)^2+\cdots+\Od\left(\delta^{4}\right),\label{eq:c4d2}\\
\left.\frac{\delta^3 {\cal L}_{\rm eff} (x)}{\delta \theta (x)^3}\right\vert_{\theta=0}=&6\sqrt 6 F v_4^{(4)}\left( \cth\, \eta'(x)-\sth\, \eta(x)\right)+\Od\left(\delta^{7/2}\right),\label{eq:c4d3}\\
\left.\frac{\delta^4 {\cal L}_{\rm eff} (x)}{\delta \theta (x)^4}\right\vert_{\theta=0}=&6 F^2 v_4^{(0)}+\Od\left(\delta^3\right),\label{eq:c4d4}
\end{align} 
where the ellipses denote further terms involving $\pi$'s and $K$'s fields, which do not contribute to connected contributions. 
The diagram displayed in Fig.~\ref{fig:c4topologies}~(a) is generated from the fourth-order derivative induced vertex in~\eqref{eq:c4d4}. 
It implies a contact term proportional to $F^2v_4^{(0)}$ and hence it only contributes at $\Od(\delta^2)$. 
The diagram (b) in Fig.~\ref{fig:c4topologies} comes from the third-order derivative induced vertex~\eqref{eq:c4d3} and the LO contribution in the $\delta$ expansion to the single-derivative vertex, i.e., first term in~\eqref{eq:c4d1}.
It involves a single $\eta^{(')}$ propagator at vanishing momentum and hence terms proportional to $F^2M_0^2 v_4^{(0)}/M^2_{0\eta^{(')}}$, where $\eta^{(')}$ stands for a $\eta$ or $\eta'$ field, that contribute at $\Od(\delta^2)$.
Diagram (c) is coming from the product of two second-order derivative vertices at NLO in the $\delta$ expansion, hence involving two $\eta^{(')}$ propagators. Nevertheless, it contributes only at $\Od(\delta^4)$ and it will not enter in the NNLO U(3) calculation for the fourth-order cumulant. The topology shown in Fig.~\ref{fig:c4topologies}~(d) is produced from the NLO U(3) contribution of the second derivative vertex in~\eqref{eq:c4d2} and two single-derivative vertices at LO in the $\delta$ expansion. 
It involves two $\eta^{(')}$ propagators and terms proportional to $M_0^4 v_2^{(2)}/M^4_{0\eta^{(')}}$ and $M_0^4 v_4^{(0)}/M^4_{0\eta^{(')}}$. Thus, it contributes at $\Od(\delta^2)$. 
Finally, diagrams (e) and (f) involve four single-derivative vertices. Diagram (e) requires one of the vertices to emit three $\eta^{(')}$ lines, i.e., the last three term in brackets in~\eqref{eq:c4d1}, 
and the three remaining producing only one meson line, which in total entails three  $\eta^{(')}$ propagators. It involves terms proportional to $M_0^6 \Lambda_2/M^6_{0\eta^{(')}}$, which are $\Od(\delta)$, and terms multiplying $M_0^6 v_2^{(2)}/M^6_{0\eta^{(')}}$,  $M_0^6 v_4^{(0)}/M^6_{0\eta^{(')}}$ and $M_0^4 \Lambda^2_2/M^6_{0\eta^{(')}}$, contributing at $\Od(\delta^2)$. Finally, at LO in U(3) diagram (f) implies an interaction vertex with four internal legs, which in turn involves four $\eta^{(')}$ propagators evaluated at zero momentum. It implies that only mass terms contribute to the interaction vertex, leading to a total contribution proportional to $F^2 M_0^8 m_{q,s}/M^8_{0\eta^{(')}}$  at $\Od(1)$, $F^2 M_0^6 \Lambda_2 m_{q,s}^2/M^8_{0\eta^{(')}}$ at  $\Od(\delta)$ or terms proportional $F^2 M_0^6 v_2^{(2)} m_{q,s}^2/M^8_{0\eta^{(')}}$ and $F^2 M_0^4 \Lambda_2^2 m_{q,s}^3/M^8_{0\eta^{(')}}$ at  $\Od(\delta^2)$. 
\begin{figure}
  \includegraphics[width=0.9\textwidth]{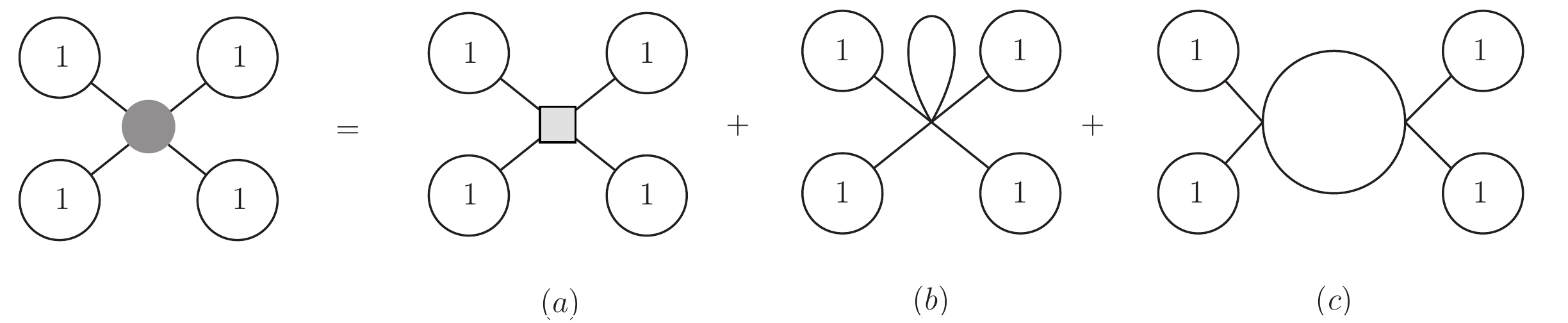}
  \caption{Topologies entering in $c_4$ from higher-order contributions to the four-point interaction vertex. Diagram (a) denotes a LECs contribution, while diagram (b) and (c) represent a tadpole or rescattering topology, respectively.}
  \label{fig:c4topologiesNLOa}
\end{figure}

Furthermore, since diagram (f) is the only one contributing at $\Od(1)$ and $\Od(\delta)$, it can be also dressed at higher orders in two different ways. On the one hand, the four-point interaction vertex can be dressed by including the higher-order diagrams depicted in~Fig.~\ref{fig:c4topologiesNLOa}. 
Namely, the LO four-point vertex can be replaced by its NLO counterpart in the U(3) expansion, i.e., by including a LEC, Fig.~\ref{fig:c4topologiesNLOa} (a).  Since the propagators are evaluated at $p=0$ this diagram only involves the LECs $L_6$, $L_7$, $L_8$ from the ${\cal L}_4$ Lagrangian and $C_{19}$, $C_{31}$ from ${\cal L}_6$.  In addition, one might add a loop to the vertex either by including a $\pi$, $K$ or $\eta^{(')}$ tadpole, Fig.~\ref{fig:c4topologiesNLOa} (b), or a rescattering diagram Fig.~\ref{fig:c4topologiesNLOa} (c). Note that in these two cases off-shell momentum-dependent terms enter. In addition, the one-loop contribution in~Fig.~\ref{fig:c4topologiesNLOa} (c) is evaluated at $p=0$, which leads to the function $M_{0i}^{2n}\log{M_{0i}^2/\mu^2}$, where $M_{0i}$ denote a $\pi$, $K$, $\eta$ or $\eta'$ and $n=0,1,2$ depending on the fields and derivatives running in the loop. 
Terms proportional to the LEC $L_8$ are $\Od{(\delta)}$. The remaining LECs and loop topologies enter only at $\Od{(\delta^2)}$.
\begin{figure}
  \includegraphics[width=0.65\textwidth]{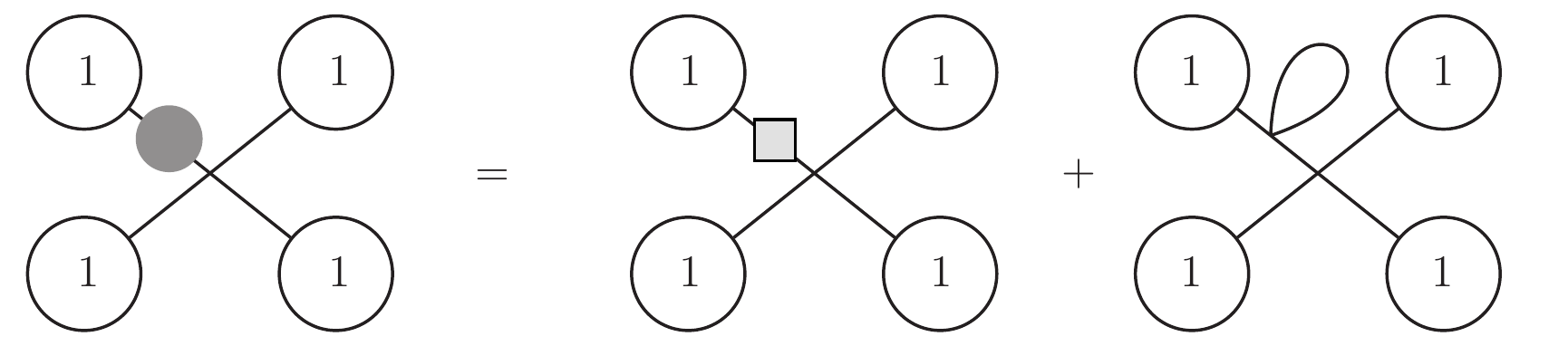}
  \caption{Topologies entering in $c_4$ from thee NLO contribution to the $\eta$ and $\eta'$ propagators. 
}
  \label{fig:c4topologiesNLOb}
\end{figure}
On the other hand, one or more of the $\eta^{(')}$ propagators can be dressed at NLO in the U(3) expansion. The $\eta^{(')}$ self energies were computed in~\cite{Guo:2015xva} in U(3) ChPT. At zero momentum they involve the LECs $\Lambda_2$, $L_8$ at  $\Od{(\delta)}$ and $L_6$, $L_7$, $L_{25}$, $v_2^{2}$, $L_8^2$, $L_8 \Lambda_2$, $\Lambda^2_2$ at $\Od{(\delta^2)}$, Fig.~\ref{fig:c4topologiesNLOb} (a), as well as further $\pi$, $K$ or $\eta^{(')}$ tadpole diagrams entering at $\Od{(\delta^2)}$, Fig.~\ref{fig:c4topologiesNLOb} (b). Finally, the product of the topology in Fig.~\ref{fig:c4topologiesNLOa} (a) proportional to $L_8$ and in Fig.~\ref{fig:c4topologiesNLOb} (a) multiplying $\Lambda_2$ and $L_8$ also contribute at  $\Od{(\delta^2)}$. 

All together, the LO contribution in the $\delta$ expansion to the fourth-order cumulant in the isospin limit reads
\begin{align}
c_4^{U(3),\, LO, \,IL }&=F^2M_0^8B_0\Bigg[\frac{\sth^4}{162 M_{0\eta}^8}\left(m_q\left(\cth^4-4\sqrt2\cth^3\sth+12\cth^2\sth^2-8\sqrt2\cth\sth^3+4\sth^4\right)\right.\nt\\
&\quad\left.+2m_s\left(4\cth^4+8\sqrt2\cth^3\sth+12\cth^2\sth^2+4\sqrt2\cth\sth^3+\sth^4\right)\right)\nt\\
&\quad-\frac{2\cth\sth^3}{81 M_{0\eta}^6 M_{0\eta'}^2}\left(m_q\left(\sqrt 2\cth^4-5\cth^3\sth+3\sqrt 2\cth^2\sth^2+2\cth\sth^3-2\sqrt2\sth^4\right)\right.\nt\\
&\quad\left.-2m_s\left(2\sqrt 2\cth^4+2\cth^3\sth-3\sqrt 2\cth^2\sth^2-5\cth\sth^3-\sqrt2\sth^4\right)\right)\nt\\
&\quad+\frac{\cth^2\sth^2(m_q+2m_s)}{27 M_{0\eta}^4 M_{0\eta'}^4}\left(2\cth^4-2\sqrt2\cth^3\sth-3\cth^2\sth^2+2\sqrt2\cth\sth^3+2\sth^4\right)\nt\\
&\quad+\frac{2\cth^3\sth}{81 M_{0\eta}^2 M_{0\eta'}^6}\left(m_q\left(-2\sqrt 2\cth^4-2\cth^3\sth+3\sqrt 2\cth^2\sth^2+5\cth\sth^3+\sqrt2\sth^4\right)\right.\nt\\
&\quad\left.+2m_s\left(\sqrt 2\cth^4-5\cth^3\sth+3\sqrt 2\cth^2\sth^2+2\cth\sth^3-2\sqrt2\sth^4\right)\right)\nt\\
&\quad+\frac{\cth^4}{162 M_{0\eta'}^8}\left(m_q\left(4\cth^4+8\sqrt2\cth^3\sth+12\cth^2\sth^2+4\sqrt2\cth\sth^3+\sth^4\right)\right.\nt\\
&\quad\left.+2m_s\left(\cth^4-4\sqrt2\cth^3\sth+12\cth^2\sth^2-8\sqrt2\cth\sth^3+4\sth^4\right)\right)\Bigg].
\label{eq:c4LOILgen}
\end{align}

Expressing once more the tree-level $\eta$ and $\eta'$ masses and mixing angle in terms of $M_0$, $m_q$ and $m_s$,~\eqref{eq:c4LOILgen} simplifies remarkably leading to
\begin{equation}\label{eq:c4LOIL}
c_4^{U(3),\, LO, \,IL }=-\frac{B_0 F^2 M_0^8 m_q m_s(m_q^3 + 2 m_s^3))}{(M_0^2(m_q+ 2m_s) + 6B_0 m_qm_s)^4}=-\frac{\Sigma \bar m_{IL}^4}{{\bar m}^{[3]}_{IL}\left(M_0^2+6B_0\bar m_{IL}\right)^4}=-\frac{\Sigma \hat m_{IL}^4}{{\bar m}^{[3]}_{IL}}, 
\end{equation}
 where $\bar m_{IL}$ and $\hat m_{IL}$ were defined in~\eqref{eq:barm}~and~\eqref{eq:hatm}, respectively, and 
\begin{equation}
{\bar m}^{[n]}_{IL}= \left[\frac{2}{m_q^n}+\frac{1}{m_s^n}\right]^{-1}.
\label{mbarndef}
\end{equation}

Like the topological susceptibility, $c_4$ vanishes in the chiral limit $m_q\to 0$ pointing out once more that it is a chiral quantity. The SU(3) result for the fourth-order cumulant at LO in the chiral expansion can be recovered by taking the limit $M_0\to\infty$. In that case $\hat m_{IL}\to \bar m_{IL}$ and one retrieves the results in~\cite{Mao:2009sy,Bernard:2012ci}.
In addition, one can also study the opposite limit, i.e., the quenched approximation for which $M_0<<\hat m$. In that case one obtains
\begin{equation}
c_4^{U(3),\,LO,\,IL}= -\frac{F^2M_0^8}{1296 B_0^3 \bar m_{IL}^{[3]}}+\Od{\left(\frac{1}{N_c^4}\right)}.
\end{equation}  

The NLO $\Od{(\delta)}$ results are always proportional to the $\Od(N_c)$ LECs $\Lambda_2$ and $L_8$, while NNLO $\Od{(\delta^2)}$ results involve the remaining pieces. Namely, terms proportional to the LECs $L_6$, $L_7$, $L_{25}$, $C_{29}$, $C_{31}$, $v_4^{(0)}$, $v_2^{(2)}$, $L_8^2$,  $L_8 \Lambda_2$, $\Lambda^2_2$ and meson logarithms. 
The renormalization of the LECs \cite{HerreraSiklody:1996pm,Kaiser:2000gs, Guo:2015xva} render a finite and scale-independent result. The explicit formulas are too long to be displayed here and they are provided as supplementary material.
  In addition, in the SU(3) $M_0\to\infty$ limit we recover the NLO results in the chiral expansion given in~\cite{Bernard:2012ci}. 

Finally, we can also study the large-$N_c$ expansion of the fourth-order cumulant. While the LO $N_c$ behavior of $\chi_{top}$ was well established time ago~\cite{Witten:1979vv,Veneziano:1979ec,Leutwyler:1992yt}, the large-$N_c$ behavior of $c_4$ is still under debate. On the one hand, in~\cite{Vicari:2008jw,Bonati:2016tvi} it was argued from the large-$N_c$ structure of the vacuum energy density that the fourth-order cumulant should scale as $\Od(1/N_c^{2})$. On the other hand, an explicit calculation  based on $U(N_f)$ ChPT at NLO in the chiral expansion for degenerate quark masses~\cite{Vonk:2019kwv} suggests that it goes as $\Od(1/N_c^{3})$.
Our LO  $\Od(1)$ and NLO  $\Od(\delta)$ results do indeed reproduce the predictions in~\cite{Vonk:2019kwv}, since we obtain that they are $\Od(1/N_c^{3})$ and $\Od(1/N_c^{4})$, respectively. 
Nevertheless, at NNLO $\Od(\delta^2)$ the fourth-derivative contribution in~\eqref{eq:c4d4} involves the contact term $-6 F^2v_4^{0}\sim\Od(1/N_c^2)$. In fact, one can show that it is the only term contributing to $c_4$ at this large-$N_c$ order. Namely, the leading-$N_c$ dependence of any $U(N_f)$ chiral operator $\hat O(X)$ involving the field X in~\eqref{Xfield} is given by~\cite{Leutwyler:1996np,HerreraSiklody:1996pm}
\begin{equation}
\hat O(X)= N_c^{2-\#(Tr)-\#(X)}, 
\end{equation}
where $\#(Tr)$ and $\#(X)$ denote the number of chiral flavor traces and powers of the operator X in~\eqref{Xfield}, respectively.
Thus, any higher order operators in $X$ will be suppressed both by their large-$N_c$ counting  and the pion decay constants $F$. 
It also implies that any $n$th-order cumulant of the topological charge distribution should scale as~$N_c^{2-n}$. All together, we obtain
\begin{align}
c_4^{U(3),\,IL}=&-6F^2 v_4^{0}-\frac{F^2M_0^8}{1296B_0^3\bar m^{[3]}_{IL}}
-\frac{4F^2M_0^2}{\bar m_{IL}B_0}\left[v_4^{0}+\frac{M_0^2}{216}\left(v_2^{(2)}-\Lambda_2^2\right)+\frac{M_0^6}{27F^4}\left(3C_{19}^r+4C_{31}^r+\frac{160}{3}L_8^{r\,2}\right)\right]\nonumber\\&+\Od{\left(\frac{1}{N_c^4}\right)} + \Od(\delta^3).
\end{align}
where, as explained, the first term in the r.h.s. is $\Od\left(1/N_c^2\right)$ and the rest of the displayed terms are $\Od\left(1/N_c^3\right)$. 

\section{Numerical results}\label{sec:num}

In Table~\ref{tab:num} we provide the numerical results for $\chi_{top}^{1/4}$, $(-c_4)^{1/4}$ and $b_2=c_4/(12\chi_{top})$ (the latter is defined following standard lattice analyses, see below) calculated in ChPT for SU(2), SU(3) and U(3) at different orders in the chiral or $\delta$ expansion. The numerical values for  the parameters involved, i.e., $F$, $M_0$, meson masses and the LECs $L_6^r$, $L_7^r$, $L_8^r$, $\Lambda_1$, $\Lambda_2$, $C_{19}$ and $C_{31}$  are taken from~\cite{Guo:2015xva}. Note that the U(3) LECs do not correspond to the usual SU(3) ChPT quantities, but larger differences might be expected between them.  More precisely, we consider the values of the NNLOFit-B, i.e., their best fit to lattice results for the $\eta$ and $\eta'$ masses. However, the constants $L_{25}^r$, $v_4^{(0)}$ and $v_2^{(2)}$ are not included in~\cite{Guo:2015xva}. $L_{25}^r$ and $v_2^{(2)}$ were estimated in~\cite{Gu:2018swy} in an additional fit to lattice data for the $\eta$ and $\eta'$ parameters and they enter both in $\chi_{top}^{1/4}$ and $c_4$. 
On the contrary, $v_4^{(0)}$ remains unknown and it only contributes to the fourth-order cumulant. 
In addition, while these LECs play a very small role on the topological susceptibility, $v_2^{(2)}$ and $v_4^{(0)}$ have a much more sizable effect on $c_4$.
In fact, taking  $L_{25}^r$ and $v_2^{(2)}$ from~\cite{Gu:2018swy} and assuming the NNLO U(3) correction for $c_4$ to agree within uncertainties with the NLO estimate, one obtains for $v_4^{(0)}=218(10)$.  Nevertheless, in order to avoid any bias in the behavior of the U(3) expansion, we will set instead te value of $v_4^{(0)}$ value to zero.
Thus, since there are no current estimates for $v_4^{(0)}$ and it would be inconsistent to include only $v_2^{(2)}$, we simple set the values of $L_{25}^r$, $v_4^{(0)}$ and $v_2^{(2)}$ to zero. 
Furthermore, we will neglect the NNLO U(3) corrections coming from $\Lambda_2^2$ since they are of the same other than the  $v_2^{(2)}$ and $v_4^{(0)}$ effects that we are ignoring. 

We also include the LECs uncertainties quoted in~\cite{Guo:2015xva}. One can see in Table~\ref{tab:num} that the uncertainties of the U(3) LECs are much larger than the standard SU(2) and SU(3) errors, the main source of error coming from $F^2$.  In that sense, 
let us remark that the SU(3) and SU(2) values quoted in Table~\ref{tab:num} are obtained from the U(3) expressions by taking the $M_0\rightarrow\infty$ limit, plus the $m_s\rightarrow\infty$ one in SU(2), but keeping the numerical values of the U(3) LECs and their uncertainties.
The reason for this is that our main purpose here is  to calibrate the numerical effect of the $\eta'$ as compared with the rest of the light degrees of freedom. Recall that, according to our previous discussion, the NNLO U(3) contribution includes the NLO SU(3) one in the limit where the $\eta'$ is decoupled and the NLO SU(2) results when also kaons and eta decouple.  This has however the drawback of losing numerical precision with respect to the corresponding  purely SU(2) and SU(3) LECs.
For instance, recent estimates of the pure SU(2) calculation are $\chi_{top}^{1/4}=75.5(5)$ MeV at NLO ~\cite{diCortona:2015ldu} and $\chi_{top}^{1/4}=75.44(34)$ MeV at NNLO~\cite{Gorghetto:2018ocs} for typical LECs values of SU(2) and SU(3)~\cite{Aoki:2019cca}. Note that the NNLO calculated for $SU(2)$ in~\cite{diCortona:2015ldu}   includes ${\cal O}(p^6)={\cal O}(\delta^3,\,\delta^4)$ corrections, which are beyond our present analysis. In particular, we do not recover those corrections when taking the $M_0\rightarrow  \infty$ and $m_s\rightarrow\infty$ limits. For that reason, we do not include those results in Table~\ref{tab:num} where, as explained, we aim to discuss the different contributions that are obtained as limiting cases of our present approach. In any case, it is pointed out in~\cite{diCortona:2015ldu}, despite the uncertainties in the ${\cal O} (p^6)$ LECs, the NNLO SU(2) numerical corrections are one order of magnitude smaller than the NLO ones.  In the case of the $b_2$ coefficient, the LO SU(2) and SU(3) expressions are LEC independent and therefore, they are given without theoretical uncertainty in Table~\ref{tab:num}. The cancellation of the $F^2$  dependence  in the U(3) LO for $b_2$, e.g. from \eqref{chitopu3lob} and \eqref{eq:c4LOIL},  explains also its smaller error compared to higher orders. 

The results in Table~\ref{tab:num} are obtained in the isospin limit. In the case of the topological susceptibility they have to be compared with the lattice result $\left[\chi_{top}^{latt}\right]^{1/4}=73(9)$ MeV in that case~\cite{Bonati:2015vqz}. As for lattice results on $c_4$,  customarily given in terms of the $b_2$ coefficient,   they are provided only for pure gauge $SU(N)$ theories in \cite{Vicari:2008jw,Panagopoulos:2011rb,Bonati:2016tvi} and for domain-wall $N_f=2+1$ fermions with large light quark masses  $m_l/m_s\geq 0.25$ \cite{ Chiu:2008jq}. Remarkably, the value of $b_2$ seems to be quite stable under those different approximations and close to the simplest SU(2) ChPT value, as pointed out in \cite{Bonati:2015vqz}. We quote for reference the isospin-limit value for $N_c=3$ gluodynamics, $b_2=-0.0216(15)$ \cite{Bonati:2016tvi}.  Nevertheless, more accurate lattice determinations for  $c_4$ and $b_2$ for the physical $N_f=2+1$ case would be needed to make further claims.

\begin{table}[h!]
  \renewcommand{\arraystretch}{1.3}
  \centering
  \begin{tabular}{|l|lll|}
    \hline
    \multicolumn{1}{|c|}{$\chi_{top}^{1/4}$ [MeV]} 
			&
			\multicolumn{1}{c}{U(3)}&
			\multicolumn{1}{c}{SU(2)}&
			\multicolumn{1}{c|}{SU(3)} \\
			\hline
			LO& 74(3) & 75(3) & 75(3) \\
			NLO& 74(3) & 78(3) & 83(2)\\
			NNLO& 81(2) & &\\
    \hline
    {\bf Lattice} &\multicolumn{3}{c|}{73(9)~\cite{Bonati:2015vqz}}\\
    \hline
		\end{tabular}
                \quad
                \renewcommand{\arraystretch}{1.3}                
               \begin{tabular}{llll|}
			\hline
			\multicolumn{1}{|c|}{$(-c_4)^{1/4}$ [MeV]} 
			&
			\multicolumn{1}{c}{U(3)}&
			\multicolumn{1}{c}{SU(2)}&
			\multicolumn{1}{c|}{SU(3)}  \\
			\hline
			\multicolumn{1}{|c|}{LO}& 50(3) & 53(2) & 52(2) \\
			\multicolumn{1}{|c|}{NLO}& 50(3) & 60(2) & 61(2) \\
			\multicolumn{1}{|c|}{NNLO}& 58(2) &&\\
                 \hline
                 \newline
               \end{tabular}
                 		\quad
		\vspace*{0.2cm}
                \renewcommand{\arraystretch}{1.3}		
                \begin{tabular}{|l|lll|}
			\hline
			\multicolumn{1}{|c|}{$b_2=\frac{c_4}{12\chi_{top}}$} 
			&
			\multicolumn{1}{c}{U(3)}&
			\multicolumn{1}{c}{SU(2)}&
			\multicolumn{1}{c|}{SU(3)}  \\
			\hline
			LO& -0.01737(4) & -0.02083 & -0.01960 \\
			NLO& -0.018(2) & -0.029(2) &  -0.025(1)\\
                  NNLO& -0.023(2) &&  \\
                  \hline
                  {\bf Lattice}&\multicolumn{3}{c|}{-0.0216(15)~\cite{Bonati:2016tvi}}\\
			\hline
		\end{tabular}
	\caption{Topological susceptibility, fourth-order cumulant  and the $b_2$ coefficient, calculated in SU(2), SU(3) and U(3) ChPT to LO, NLO and NNLO in the isospin limit. The lattice values mentioned in the main text have been also quoted here for reference. The numerical values of the masses, decay constants and LECs involved, as well as their uncertainties, are taken from~\cite{Guo:2015xva}, except  $L_{25}$, $v_2^{(0)}$ and $v_4^{(0)}$, which are set to zero, (see main text).}
	\label{tab:num}
	\end{table}

From the results in Table~\ref{tab:num} we also observe the following features.  First, in the three theoretical frameworks, the perturbative corrections remain reasonably under control. Second, although the SU(2) approach already reproduces the main contribution, the $\eta'$ meson and mixing angle corrections that we are including in the present work are actually comparable to the kaon and $\eta$ ones introduced in the SU(3) approach. For $\chi_{top}$, those corrections lower the central value and get closer to the lattice prediction and so on for $\vert b_2\vert$. Actually,  we  see that the full U(3) calculation for both observables remains compatible with the lattice results within the range provided by the LECs uncertainties, which in the case of $\chi_{top}$ holds also for all the different approximations collected in Table~\ref{tab:num}. The latter confirms that these are are good chiral quantities in the sense that they can be accurately described within ChPT.

\section{isospin breaking corrections to $\chi_{top}$ and $c_4$}
\label{sec:ib}

As mentioned in the introduction, isospin breaking corrections can become important for the topological charge distribution.
The main reason is that for $\theta\neq 0$ and $m_u\neq m_d$, the constant field configuration that minimizes the vacuum energy density is not $U_0=\ID$, but generally $U_0=\diag \left(e^{\phi_1},\dots,e^{\phi_{N_f}}\right)$ ~\cite{DiVecchia:1980yfw,Gasser:1984gg,Mao:2009sy,Guo:2015oxa}.
In the SU(3) framework, the constraint $\det U_0=1$ leads to $\sum_j \phi_j=0$.
However, within our present U(3) formalism, such constraint does not hold, since the determinant of $U_0$ is an additional degree of freedom~\cite{DiVecchia:1980yfw}. The $U_0$ configuration should be such that the vacuum energy density $\epsilon_{vac} (\theta)$ defined in~\eqref{evacdef} is minimized,
which is indeed achieved for a constant value of $U_0$.
Therefore, to LO in the chiral expansion, we have  to consider just the usual NGB mass term plus the $M_0^2$ term in~\eqref{efflag0} for the Euclidean action. Namely,
\begin{eqnarray}
\epsilon_{vac}^{LO}(\theta)&=&-\frac{F^2 B_0}{2}\tr\left[U_0{\cal M}^\dagger+{\cal M} U_0^\dagger\right]-\frac{F^2}{12}M_0^2\left[i\theta+\log\det U_0\right]^2
\nonumber\\
&=&-F^2 B_0 \sum_{j=1}^{N_f} m_j\cos\left[\phi_j (\theta)\right]+\frac{F^2 M_0^2}{12}\left(\theta+\sum_{j=1}^{N_f}\phi_j (\theta)\right)^2\ec
\label{epsilonloib}
\end{eqnarray}
with ${\cal M}=\diag (m_u,m_d,m_s)$ the quark mass matrix and $\phi_j(\theta)$ are such that they minimize $\epsilon_{vac}^{LO}(\theta)$, i.e., 
\begin{equation}
B_0 m_k\sin\left[\phi_k(\theta)\right] +\frac{M_0^2}{6}\left(\theta+\sum_{j=1}^{N_f} \phi_j (\theta)\right)=0 \qquad (k=1,\dots,N_f)\ep
\label{min}
\end{equation}

Note that the solution to~\eqref{min} is equivalent to encode the $\theta$ dependence in a complex quark mass matrix ${\cal M}\exp(-i\theta/N_f)$ as in~\cite{Mao:2009sy,Guo:2015oxa} with the change of variable $\phi_j\rightarrow \phi_j +\theta/N_f$. 

Now, following the same procedure as in~\cite{Mao:2009sy}, in order to solve the minimization problem we expand $\cos\phi_j$ in powers of $\phi_j$. The reason for this is that we are only interested in the power expansion of $\epsilon_{vac}(\theta)$ around $\theta=0$ and hence around the solution $\phi_j=0$ of ~\eqref{min} for $\theta=0$. For $N_f=2$ and $M_0=0$ the solution of~\eqref{min} can be found in~\cite{Mao:2009sy,Guo:2015oxa}. The solution of the system~\eqref{min} to $\Od(\phi_j)$ for $N_f=3$ is  $\phi_j(\theta)=\phi_j^0 \theta+\Od(\theta^3)$ with
\begin{equation}
\phi_j^0=-m_j \left(\frac{6B_0}{M_0^2}+\frac{1}{\bar m}\right)  \qquad (j=u,d,s)\ec
\label{solphilinear}
\end{equation}
with $\bar m$ defined in~\eqref{chitopsu3lo}. Replacing the above linear order in the vacuum energy density~\eqref{epsilonloib} yields
\begin{equation}
\epsilon_{vac}^{LO}(\theta)=\epsilon_{vac}(0)+\frac{1}{2}\chi_{top}^{U(3),LO} \theta^2 + \Od(\theta^4)\ec
\label{energy2}
\end{equation}
with
\begin{equation}
\chi_{top}^{U(3),LO}= \Sigma {\hat m}\ec
\label{chitoploib}
\end{equation}
and
$$
{\hat m}=\frac{M_0^2 {\bar m}}{M_0^2+6B_0{\bar m}}\ec
$$

 The above result is the extension to $m_u\neq m_d$ of the LO U(3) result in~\eqref{chitopu3lob}, which amounts to the replacement ${\bar m}_{IL}\rightarrow {\bar m}$ and reproduces the LO result with isospin breaking in SU(3) in~\eqref{chitopsu3lo} by replacing ${\bar m}\rightarrow {\hat m}$. 

In order to provide a numerical estimate of the isospin breaking effect, we write~\eqref{chitoploib} as
\begin{equation}
\chi_{top}^{U(3),LO}=\frac{F^2 M_0^2}{6+\frac{(1+z)^2}{z}\frac{M_0^2}{M_{0\pi^0}^2}+\frac{(1+z)M_0^2}{(1+z)M_{0 K^0}^2-M_{0\pi^0}^2}}
\label{chitoploib2}
\end{equation}
where $z=m_u/m_d$. Using the central value $z=0.485$ of the recent lattice analysis~\cite{Aoki:2019cca}, we get $\left[\chi_{top}^{U(3),LO}\right]^{1/4}=72$ MeV, to be compared with the U(3) LO value in Table \ref{tab:num}, which corresponds to $z=1$ in~\eqref{chitoploib2}. Thus, the isospin correction to LO U(3) is within the 5\% range and lies within the theoretical LO uncertainty. It is therefore numerically safe to consider isospin breaking only for the LO U(3) result in our present analysis.

Following the same approach to the next order in the $\theta$ expansion allows us to calculate the isospin breaking corrections to the fourth-order cumulant. Thus, we expand~\eqref{min} up to $\Od(\phi_j^3)$ and write its solution as $\phi_j(\theta)=\phi_j^0\theta + \phi_j^1\theta^3 + \Od(\theta^5)$ and keep only up to $\Od(\theta^3)$ in the equation, thus solving linearly for the  $\phi_j^1$. Replacing then the solution in~\eqref{epsilonloib} yields:

\begin{equation}
\epsilon_{vac}^{LO}(\theta)=\epsilon_{vac}(0)+\frac{1}{2}\chi_{top}^{U(3),LO} \theta^2 + \frac{1}{24} c_4^{U(3),LO} \theta^4+\Od(\theta^6)
\label{energy4}
\end{equation}
with the fourth order cumulant

\begin{equation}
c_4^{U(3),LO}=-\Sigma \frac{{\hat m}^4}{{\bar m}^{[3]}}
\label{cumulantloib}
\end{equation}
where, following the notation of~\cite{Bernard:2012ci}, we have defined, consistently with \eqref{mbarndef},

\begin{equation}
{\bar m}^{[3]}= \left[\frac{1}{m_u^3}+\frac{1}{m_d^3}+\frac{1}{m_s^3}\right]^{-1}.
\end{equation}

The result~\eqref{cumulantloib} for the fourth-order cumulant corresponds again to the SU(3) one in~\cite{Mao:2009sy,Bernard:2012ci,Guo:2015oxa} with the replacement $\bar m \rightarrow \hat m$. As in the previous case, we also can write in terms of measurable meson parameters:

\begin{equation}
c_4^{U(3),LO}=-\frac{F^2 M_0^8(1+z)^3 \left[\frac{1}{\left[(1+z) M_{0K^0}^2-M_{0\pi^0}^2\right]^3}+\frac{1+z^3}{z^3 M_{0\pi^0}^6}\right]}{\left[6+\frac{(1+z)M_0^2}{(1+z) M_{0K^0}^2-M_{0\pi^0}^2}+\frac{(1+z)^2}{z}\frac{M_0^2}{M_{0\pi^0}^2}\right]^4}
\label{cumulantloib2}
\end{equation}
which yields $(-c_4^{U(3),LO})^{1/4}$=53 MeV, so that the correction lies also within the 5\% level when compared with the isospin-limit values in Table \ref{tab:num}.

Finally, we recall that at the order we are considering the isospin-breaking corrections, temperature dependence  is absent since it only enters through loop contributions. 

\section{Finite temperature dependence}
\label{sec:temp}

From our present U(3) ChPT analysis, we can straightforwardly include the temperature dependence coming from  meson loops. In the case of $\chi_{top}$, loop effects only arise at NNLO in the $\delta$ expansion from tadpole contributions coming from the Euclidean tree-level propagator $G_i(x=0)$. Its finite part reads
\begin{align}
\mu_i(T)=\frac{M_{0i}^2}{32\pi^2 F^2}\log\frac{M_{0i}^2}{\mu^2}+\frac{g_1(M_{0i},T)}{2F^2}\ec \\
g_1(M,T)=\frac{T^2}{2\pi^2}\int_{M/T}^\infty \diff x  \frac{\sqrt{x^2-(M/T)^2}}{e^{x}-1}\ec
\label{mudef}
\end{align}     
where  $i=\pi,K,\eta,\eta'$, $M_{0i}$ are the tree level masses and $\mu$ is the  renormalization scale.

In the case of $c_4$, in addition to tadpoles, which enter again from the Euclidean tree-level propagator $G_i(x=0)$ in Fig~\ref{fig:c4topologiesNLOb} (b) but also from the six-point interaction vertex in Fig.~\ref{fig:c4topologiesNLOa} (b), one has to take into account the  one-loop function depicted in Fig.~\ref{fig:c4topologiesNLOa} (c), which finite part  can be written in terms of
\begin{align}
\nu_i (T)=&F^2\frac{d}{d M_{0i}^2}  \mu_i(T) = \frac{1}{32\pi^2} \left[1+\log\frac{M_{0i}^2}{\mu^2}\right]-\frac{g_2(M_{0i},T)}{2},\\
g_2(M,T)=&\frac{1}{4\pi^2}\int_{M/T}^\infty dx  \ \frac{1}{x}\frac{1}{e^{x}-1}.
\label{nudef}
\end{align}

\begin{figure}
	\centering
	\subfloat{
		\includegraphics[width=9cm]{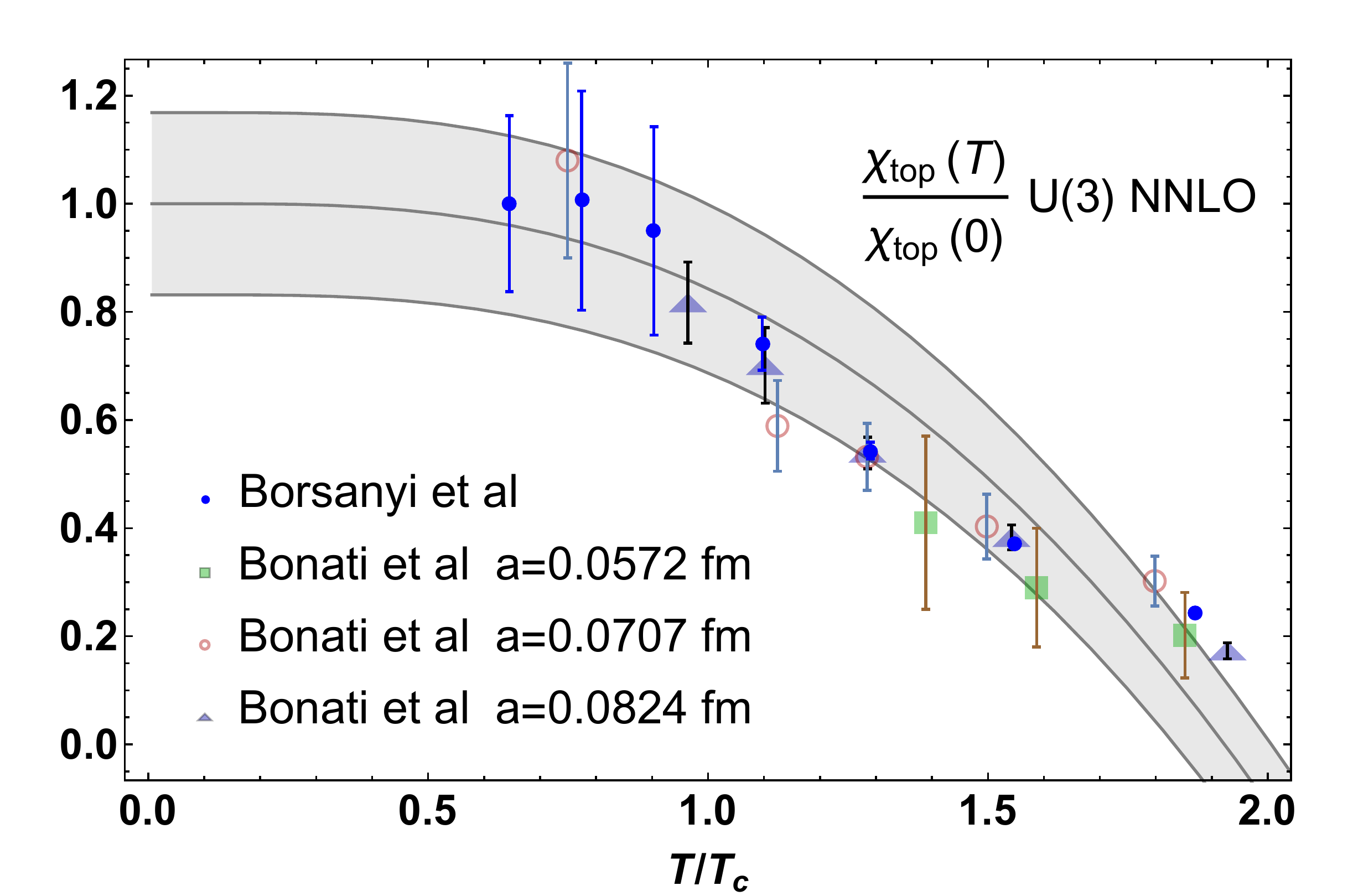}	\includegraphics[width=9cm]{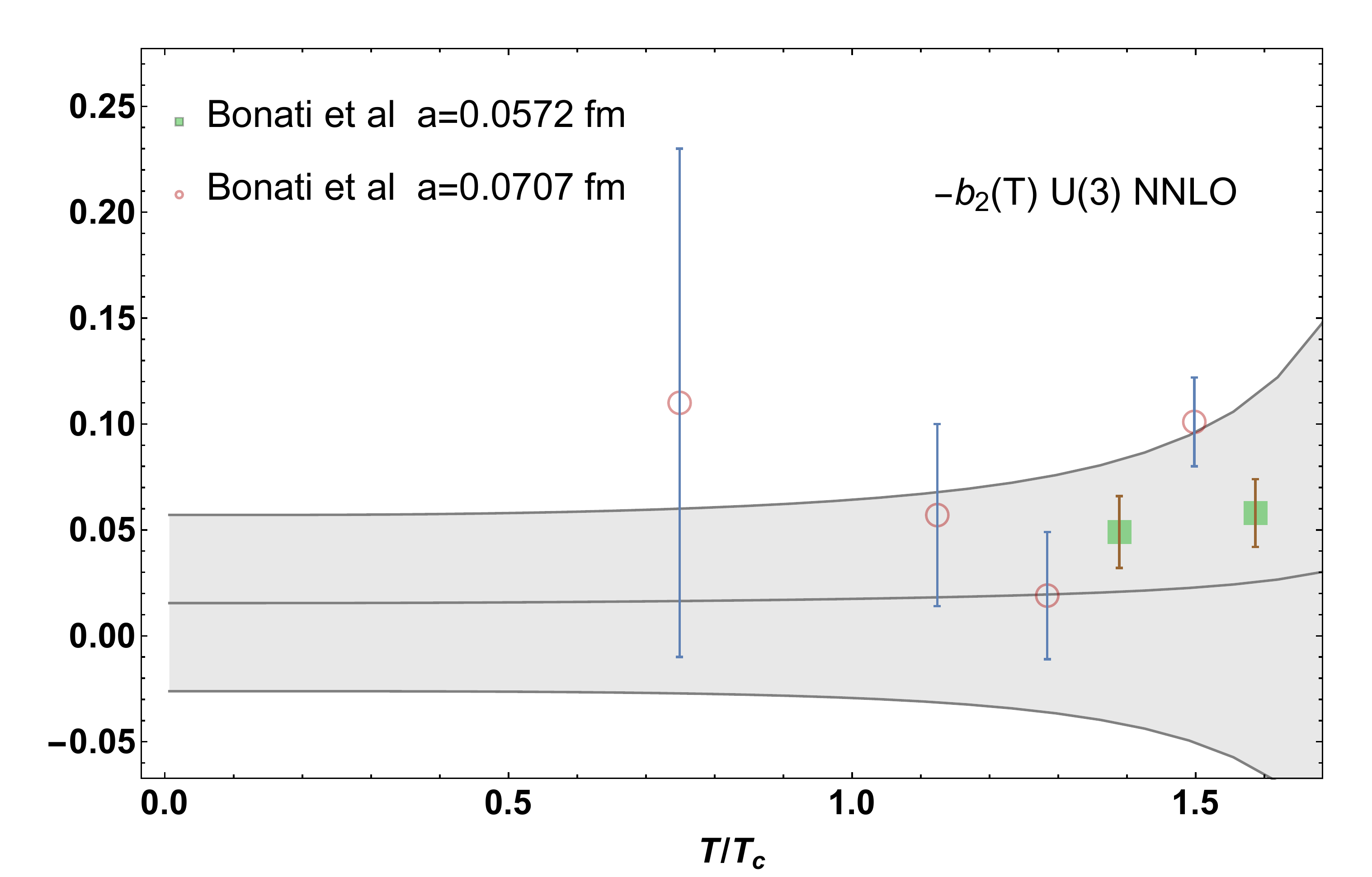}}
	\caption{Left: Temperature dependence of the topological susceptibility calculated within the U(3) formalism compared to lattice data from ~\cite{Bonati:2015vqz} and~\cite{Borsanyi:2016ksw}. Here, we have taken $T_c=155$ MeV. Right: Temperature dependence of the $b_2$ coefficient, with lattice data from~\cite{Bonati:2015vqz}.}
	\label{fig:temp1}
\end{figure}

That said, we want to remark that the U(3) thermal expansion is based on a perturbative calculation and hence, it is only expected to converge at very low temperatures.
The fact that thermal corrections arise at NNLO in the $\delta$ expansion implies that the series breaks down as soon as thermal effects are sizable. 
In Figure~\ref{fig:temp1}, we  show the temperature dependence of $\chi_{top}(T)$ and $b_2(T)$ for the same parameter values and errors used for the $T=0$ results in Table \ref{tab:num}.  In addition, we also plot the lattice data results for the topological susceptibility obtained in~\cite{Bonati:2015vqz} and~\cite{Borsanyi:2016ksw} and for $b_2$ in~\cite{Bonati:2015vqz}. In the latter case, lattice errors are  larger than for the susceptibility and we have actually not considered the data set for $a=0.0824$ fm which has even  larger errors than those showed in the figure. 

We see that the present U(3) ChPT analysis is consistent with the lattice within uncertainties, even beyond its applicability range, which as mentioned before lies at low temperatures, well below the transition. Consequently, its extrapolation close and above $T_c$ has to be taken with care.  
However, the good agreement with the lattice observed in Figure~\ref{fig:temp1} reveals once more that accounting properly for the lightest meson degrees of freedom is crucial for the description of the topological susceptibility and the cumulant.

\begin{figure}
	\centerline
	{\includegraphics[width=9cm]{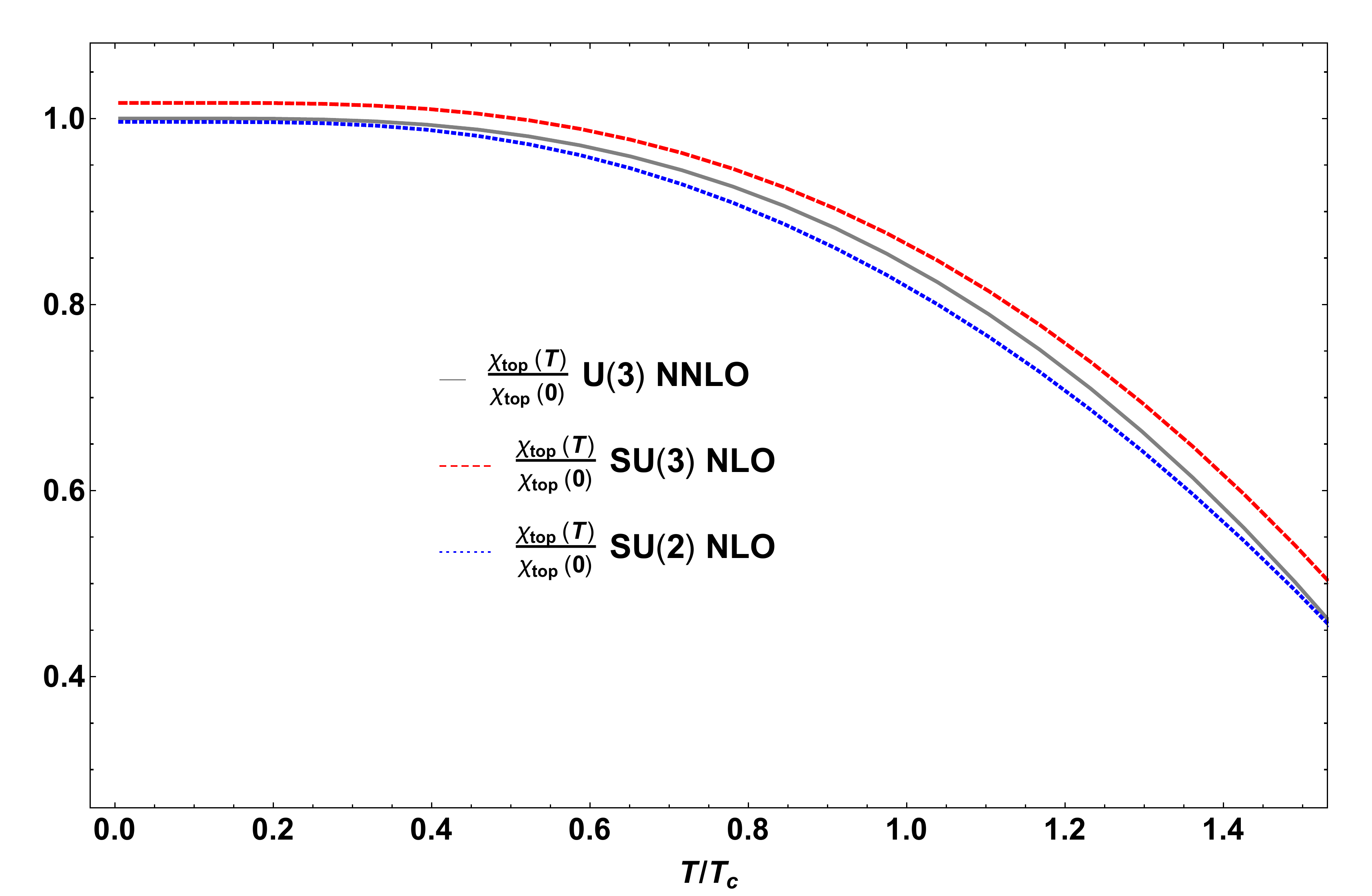} \includegraphics[width=9cm]{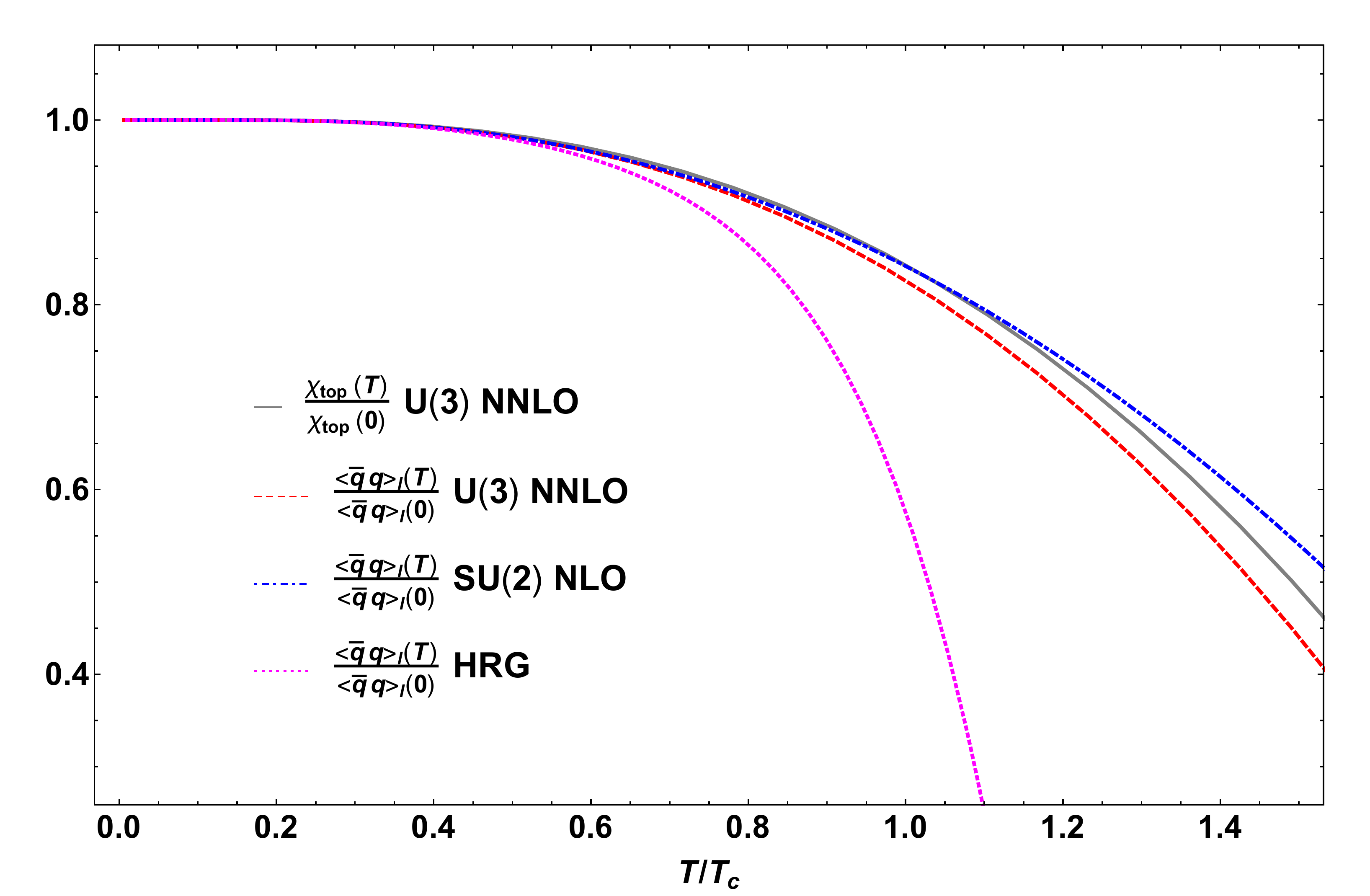}}
	\caption{Left: comparison of  the U(3), SU(3) and SU(2) limits for the topological susceptibility. Right: comparison of the topological susceptibility scaling at finite temperature with various approaches of the light quark condensate scaling. }
	\label{fig:temp2}
\end{figure}

This implies an important difference with other thermodynamic observables like the quark condensate, which accurate description based on effective theories requires the contribution of many hadronic states,
like for instance in the Hadron Resonance Gas approach~\cite{Karsch:2003vd,Tawfik:2005qh,Huovinen:2009yb,Jankowski:2012ms}.   
Actually, in figure \ref{fig:temp2}  we compare different orders of the finite-$T$ ChPT approach for the topological susceptibility and we see that the SU(3) and U(3) calculations represent rather small deviations from the SU(2) one. 

In the case of $b_2$, our theoretical $U(3)$ result in Figure~\ref{fig:temp1} is quite flat with temperature, and lattice data are less accurate. This indicates  that both quantities $\chi(T)$ and $c_4(T)$ decrease with $T$ in a roughly similar way. Nevertheless, the agreement between theory and lattice results is also quite remarkable. 

As discussed in section \ref{sec:intro}, another important issue regarding the temperature dependence is to what extent it can be approximated by just the scaling of the quark condensate, i.e.,  whether or not the second term in the  Ward Identity in~\eqref{wi1} can be ignored.  This is actually the case if one sticks to SU(2) at NLO in the chiral expansion, what has been used in~\cite{diCortona:2015ldu}. 
In our present work, we are calculating  $\chi_{top}(T)$ in U(3) at NNLO, including SU(3) and SU(2) NLO as special cases. 
Thus, we can provide a much more accurate analysis in that respect. 
That  issue can be also of relevance for lattice analyses. If the quark condensate terms dominate, the combined use of~\eqref{wi1} and~\eqref{wi2} may help to relate $\chi_{top}$ with quantities much better determined in the lattice. Namely,
 \begin{equation}
    \chi_{top}=\dfrac{m_q}{2}\left(2-\frac{m_q^2}{m_s^2}\right)^{-1}\left[2\Delta_{l,s}-m_q\left(\chi_p^{ll}-2\chi_p^{ss}\right)\right]\ec
    \label{chitopvsdeltals}
    \end{equation}
   where 
$$\Delta_{l,s}=\condl (T)-2\frac{m_q}{m_s}\conds (T)$$
is the reduced quark condensate used in lattice calculations to eliminate finite-size divergences appearing in individual condensates~\cite{Buchoff:2013nra,Bazavov:2012qja}. 
The relation~\eqref{chitopvsdeltals} offers a way to measure indirectly $\chi_{top}$, which is alternative to the usual method based on the WI 
$$\chi_{top}=\frac{m_q^2}{4}\left(\chi_P^\pi-\chi_P^{ll}\right)$$ 
which stems directly from~\eqref{wi1}, with $\chi_P^{ \pi}=-\condl/m_q$ the pion susceptibility~\cite{Nicola:2013vma}.
The reduced quark condensate is a particular example of a well determined quantity in the lattice. 

The possible dominance of the quark-condensate term in the WI~\eqref{wi1} is also relevant for a current topic of discussion, which has been actively studied both theoretically and in the lattice~\cite{Shuryak:1993ee,Cohen:1996ng,Lee:1996zy,Aoki:2012yj,Meggiolaro:2013swa,Cossu:2013uua,Buchoff:2013nra,Bhattacharya:2014ara,Azcoiti:2016zbi,Tomiya:2016jwr,Brandt:2016daq,GomezNicola:2017bhm,Nicola:2018vug,Dick:2015twa}.
Namely, whether the chiral and $U(1)_A$ restoration temperatures are close enough.  Since $\chi_{top}$ and $\condl$ are meant to vanish at exact $U(1)_A$ (asymptotically) and chiral $O(4)$ restoration, respectively, their difference, encoded in $\chi_P^{ll}$ in~\eqref{wi1}, provides a direct measure of the separation between the two transitions. 
In this sense, it is useful to recall the behavior of these quantities near the light chiral limit $m_q\rightarrow 0^+ (M_\pi\rightarrow 0^+)$, where the effects of chiral symmetry restoration are meant to be enhanced. 
In NNLO U(3) ChPT~\cite{Nicola:2018vug}, $\chi_{top}=\Od(m_q)$, while $\condl (T)$ and $\chi_P^{ll} (T)$ are both $\Od(1)$ quantities in the $m_q\rightarrow 0^+$ limit. Thus, in the equation
 \begin{equation}
\frac{\chi_{top}}{m_q}=-\frac{1}{4}\left[\condl+m_q \chi_P^{ll}\right] \label{wi1b}\ec
\end{equation}
only the quark condensate contribution survives in the right-hand side of~\eqref{wi1b} in the chiral limit, which supports its dominance at low temperatures.  
However, near the transition $\condl (T)\rightarrow 0^+$  in the chiral limit, while $\chi_P^{ll}(T)$ changes much more slowly, since it is controlled by  a term proportional to $T^2/M_{0K}^2$. 
Thus, it brings up the question as to whether near the transition the $\chi_P^{ll}$  term  can become important enough for physical masses. 
 
In Figure~\ref{fig:temp2},   we compare the temperature dependence of the full U(3) topological susceptibility with the scaling of the light quark condensate calculated in the same framework,
which would correspond to neglect the $\chi_{ll}$ term in the WI~\eqref{wi1} and~\eqref{wi1b}. 
The values of the LECs and other parameters involved are the same as those used in Figure~\ref{fig:temp1}. 
In the same figure, we also show the simple NLO SU(2) scaling used in~\cite{diCortona:2015ldu}, corresponding just to the NLO ChPT quark condensate, namely
\begin{equation}
\frac{\chi_{top}^{SU(2), NLO}(T)}{\chi_{top}^{SU(2), NLO}(0)}=\frac{\condl^{SU(2),NLO}(T)}{\condl^{SU(2),NLO}(0)}=1-\frac{3}{2F_\pi^2} g_1(M_\pi,T)\ep
\label{scalingsu2}
\end{equation}

The results in Figure~\ref{fig:temp2} show that the contribution from the additional $\chi_{ll}$ term in the WI, although not large, may be significant as $T$ approaches the transition point. 
The simple SU(2) description remains also close to the full U(3) one, which is a test of its robustness despite its simplicity. 
However, it is very important to point out that, although the topological susceptibility may be well described within a ChPT analysis  including only the lightest degrees of freedom, that is definitively not the case for the quark condensate. To show this explicitly, we have plotted in Figure~\ref{fig:temp2}  the quark condensate resulting from the HRG approximation provided in~\cite{Jankowski:2012ms}, which includes hadron states with masses up to 2 GeV and provides a very good fit to lattice condensate data. One can see clearly a much larger departure from the scaling of the topological susceptibility than the one observed with the  ChPT expressions for the quark condensate, which happen to remain close to the topological susceptibility. In this sense, we remark  that the addition of degrees of freedom is expected to reduce drastically the chiral condensate, as expected from approximate chiral restoration at $T_c$.
Nevertheless, that may not be the case for the topological susceptibility, whose behavior is not directly related to chiral symmetry restoration but includes $U(1)_A$ restoration features, describing lattice data just with the light degrees of freedom. 
Thus, in a full description of the hadron gas we do expect large deviations from the quark condensate scaling and hence significant contributions from the second term in~\eqref{wi1}, which becomes large close to the transition point.
This also indicates that the $U(1)_A$ symmetry is still sizeably broken at the chiral transition for physical quark masses. 
This analysis should prevent from the use of the topological susceptibility to extract the quark condensate at finite temperature. 

 \section{Conclusions}
 \label{sec:conc}
The main conclusions achieved in this work are the following:
\begin{itemize}
\item We have provided a full calculation of the topological susceptibility and the fourth-order cumulant up to NNLO in U(3) Chiral Perturbation Theory. Our result allows one to consider the effect of the $\eta'$ meson consistently, as well as the $\eta-\eta'$ mixing angle dependence. As limits of interest,  we recover the SU(2) and SU(3) results when $M_0\rightarrow\infty$. In addition, we  have discussed the large-$N_c$ corrections to both quantities. In the case of the topological susceptibility, we have provided the $\Od(1/N_c)$ correction to the Witten-Veneziano formula up to $\Od(\delta^2)$ in the $U(3)$ ChPT expansion, and so on for the $\Od(1/N_c^2)$ and $\Od(1/N_c^3)$ corrections to the fourth-order cumulant. 

\item  We have estimated the $\eta'$ corrections to $\chi_{top}$ and $c_4$ at zero temperature, which turn out to be of the same order as the $K$ and $\eta$ contributions. 
Furthermore, it provides results compatible with lattice analyses, consistently with the idea that the QCD topological charge is an observable well described by the expansion around the chiral limit provided by ChPT. 

\item Including the dominant isospin breaking effect in the vacuum misalignment, we have provided both $\chi_{top}$ and the fourth cumulant $c_4$ of the vacuum energy density expansion in the $\theta$ parameter to LO in the U(3) ChPT expansion. Numerical corrections due to isospin breaking remain below the 5\% level for $\chi_{top}^{1/4}$ and $(-c_4)^{1/4}$. 

\item  We have calculated $\chi_{top}$ and $c_4$ at finite temperature up to NNLO in U(3) ChPT. The temperature dependence obtained for the topological susceptibility is consistent with lattice data, supporting again that this quantity is well described by a gas made only of light mesons, unlike for instance the quark condensate. We have also discussed the relation between these two quantities, which are connected through a Ward Identity valid at all temperatures. 
Although the quark condensate calculated within the same ChPT formalism seems to scale quite similarly to $\chi_{top}$, we argue that this cannot be the case for the full hadron gas.
It reveals a sizable gap between the chiral transition and the $U(1)_A$ one for physical quark masses, even though recent theoretical analysis show that those transition tend to coincide near the chiral limit for exact chiral restoration. 
\end{itemize}
  
 \section*{Acknowledgments}
 We are very grateful to   F.~K.~Guo, U.~G.~Mei\ss ner,  J.~J.~Sanz-Cillero, Z. H. Guo and T.~Vonk for very useful comments. Work partially supported by  research contract FPA2016-75654-C2-2-P  (spanish ``Ministerio de Econom\'{\i}a y Competitividad") and the Swiss National Science Foundation, project No.\ PZ00P2\_174228. This work has also received funding from the European Union Horizon 2020 research and innovation programme under grant agreement No 824093. A. V-R acknowledges support from a fellowship of the UCM predoctoral program. 
 
 \appendix
 
 \section{Results for $\chi_{top}$ at NLO and NNLO in U(3) ChPT in the isospin limit}
 \label{app:results}
 
 We provide here the full results for the topological susceptibility in  the U(3) ChPT formalism at NLO and NNLO in the $\delta$ expansion.  
 \begin{eqnarray}
 \chi_{top}^{U(3),NLO}&=&\frac{1}{9 M_{0\eta '}^4}\left\{
 F^2 \Lambda _2 M_0^2 \cth \left[2 M_{0K}^2 \left(\cth-\sqrt{2}\sth\right)+M_{0\pi
   }^2 \left(2 \sqrt{2}\sth+\cth\right)\right]\left(M_0^2 \cth^2-M_{0\eta
   '}^2\right)\right.\nonumber\\
   &+&\left.8 M_0^4 \cth^2 L_8^r \left[4 M_{0K}^4 \left(2 \sth^2+\cth^2-2 \sqrt{2} \sth \cth\right)\right.\right.\nonumber\\
  &-&\left.\left.4 M_{0\pi }^2 M_{0K}^2 \left(2 \sth^2+\cth^2-2
   \sqrt{2}\sth \cth\right)
   +3 M_{0\pi }^4\right]\right\}\nonumber\\
   &+& \frac{1}{9 M_{0\eta}^4}\left\{F^2 \Lambda _2 M_0^4 \sth^3 \left[2 M_{0K}^2 \left(\sth+\sqrt{2} \cth\right)+M_{0\pi }^2 \left(\sth-2 \sqrt{2} \cth\right)\right]\right.
   \nonumber\\
   &+& \left.
   8 M_0^4 \sth^2
   L_8^r \left[4 M_{0K}^4 \left(\sth^2+2 \cth^2+2 \sqrt{2}\sth \cth\right)\right.\right.\nonumber\\
   &-&\left.\left.4 M_{0\pi }^2 M_{0K}^2 \left(\sth^2+2 \cth^2+2 \sqrt{2}\sth \cth\right)+3 M_{0\pi }^4\right]\right\}\nonumber\\
   &+& \frac{1}{9 M_{0\eta }^2 M_{0\eta '}^2}\left\{
   F^2 \Lambda _2 M_0^2\sth \left[
   \left(2 M_{0K}^2 \left(\sth+\sqrt{2} \cth\right)+M_{0\pi }^2 \left(\sth-2 \sqrt{2} \cth\right)\right) \left(M_0^2 \cth^2-M_{0\eta '}^2\right)\right.\right.\nonumber\\
   &+&\left.\left.M_0^2\sth \cth \left[2 M_{0K}^2 \left(\cth-\sqrt{2}\sth\right)+M_{0\pi }^2 \left(2 \sqrt{2}\sth+\cth\right)\right]\right]
   \right.\nonumber\\
  &+&\left. 64 M_0^4\sth \cth M_{0K}^2 \left(M_{0K}^2-M_{0\pi }^2\right) L_8^r \left(-\sqrt{2} \sth^2+\sqrt{2} \cth^2-\sth \cth\right)
   \right\}
   \label{chitopu3nlo}
 \end{eqnarray}

 \begin{eqnarray}
 \chi_{top}^{U(3),NNLO}&=&
 -\frac{16 M_0^2 \Lambda _2 L_8^r }{27 M_{0\eta }^6 M_{0\eta '}^6}\left\{8M_{0K}^6 \left[\cth \left(\cth^3 \sqrt{2}\sth \cth^2+6 \sth^2 \cth
    \right.\right.\right.
     \nonumber\\  
  &-&\left.\left.\left.2 \sqrt{2} \sth^3\right) \left(2 \cth^2 M_0^2-M_{0\eta '}^2\right) M_{0\eta }^6M_{0\eta '}^2 \left(2 \cth \left(\cth^5+2
   \sqrt{2}\sth \cth^4\right.\right.\right.\right.\nonumber\\
   &-&
   \left.\left.\left.\left.
 7 \sth^2 \cth^3-2 \sqrt{2} \sth^3 \cth^2+7 \sth^4 \cth-\sqrt{2} \sth^5\right) M_0^2
   \right.\right.\right.\nonumber\\
   &+&\left.\left.\left.
   \left(-2 \cth^4-\sqrt{2}\sth \cth^3+6 \sth^2 \cth^2+\sqrt{2} \sth^3 \cth-2 \sth^4\right) M_{0\eta '}^2\right) M_{0\eta }^4
   \right.\right.\nonumber\\
   &+&\left.\left.
    \sth M_{0\eta '}^4 \left(2 \left(\sqrt{2} \cth^5+7 \sth \cth^4+2 \sqrt{2} \sth^2 \cth^3-7 \sth^3 \cth^2-2 \sqrt{2} \sth^4 \cth+\sth^5\right) M_0^2
   \right.\right.\right.\nonumber\\
   &-&\left.\left.\left.
\left(2 \sqrt{2} \cth^3+6\sth \cth^2+3 \sqrt{2} \sth^2 \cth+\sth^3\right) M_{0\eta '}^2\right) M_{0\eta }^2
  \right.\right.\nonumber\\
   &+&\left.\left.
2 \sth^3\left(2 \sqrt{2} \cth^3+6\sth \cth^2+3 \sqrt{2} \sth^2\cth+\sth^3\right) M_0^2 M_{0\eta '}^6\right]
   \right.\nonumber\\
   &-&\left.
   4 M_{0\pi }^2 M_{0K}^4 \left[\cth
   \left(\cth^3-6 \sqrt{2}\sth \cth^2+18 \sth^2 \cth-8 \sqrt{2} \sth^3\right) \left(2 \cth^2 M_0^2-M_{0\eta '}^2\right) M_{0\eta }^6
   \right.\right.\nonumber\\
   &+&\left.\left.
   2
   M_{0\eta '}^2 \left(\cth \left(4 \cth^5+2 \sqrt{2}\sth \cth^4-25 \sth^2 \cth^3+4 \sqrt{2} \sth^3 \cth^2+22
   \sth^4 \cth
   \right.\right.\right.\right.\nonumber\\
   &-&\left.\left.\left.\left.
4 \sqrt{2} \sth^5\right) M_0^2+\left(-4 \cth^4+\sqrt{2}\sth \cth^3+9 \sth^2 \cth^2-\sqrt{2}
   \sth^3 \cth-4 \sth^4\right) M_{0\eta '}^2\right) M_{0\eta }^4
   \right.\right.\nonumber\\
 &+&\left.\left.
M_{0\eta '}^4 \sth\left(2 \left(4 \sqrt{2} \cth^5+22\sth \cth^4-4 \sqrt{2} \sth^2 \cth^3-25 \sth^3 \cth^2-2 \sqrt{2} \sth^4 \cth+4 \sth^5\right) M_0^2
  \right.\right.\right.\nonumber\\
 &-&\left.\left.\left.
\left(8 \sqrt{2} \cth^3+18
  \sth \cth^2+6 \sqrt{2} \sth^2 \cth+\sth^3\right) M_{0\eta '}^2\right) M_{0\eta }^2
   \right.\right.\nonumber\\
   &+&\left.\left.
2 \sth^3 \left(8 \sqrt{2} \cth^3+18 \sth \cth^2+6 \sqrt{2} \sth^2 \cth+\sth^3\right) M_0^2 M_{0\eta '}^6\right] 
   \right.\nonumber\\
   &+&\left.
2 M_{0\pi }^4 M_{0K}^2 \left[\cth \left(\cth^3-3 \sqrt{2} \sth \cth^2+15 \sth^2 \cth-11 \sqrt{2} \sth^3\right) \left(2 \cth^2 M_0^2-M_{0\eta '}^2\right) M_{0\eta }^6
  \right.\right.\nonumber\\
   &+&\left.\left.
   M_{0\eta '}^2 \left(\cth \left(8 \cth^5-5 \sqrt{2}\sth \cth^4-38 \sth^2
   \cth^3+32 \sqrt{2} \sth^3 \cth^2+38 \sth^4 \cth
   \right.\right.\right.\right.\nonumber\\
   &-&\left.\left.\left.\left.
11 \sqrt{2} \sth^5\right) M_0^2-4 \left(2 \cth^4-2 \sqrt{2} \sth \cth^3-3 \sth^2 \cth^2+2 \sqrt{2} \sth^3 \cth+2 \sth^4\right) M_{0\eta '}^2\right) M_{0\eta }^4
  \right.\right.\nonumber\\
   &+&\left.\left.
\sth M_{0\eta '}^4 \left(\left(11
   \sqrt{2} \cth^5+38\sth \cth^4-32 \sqrt{2} \sth^2 \cth^3-38 \sth^3 \cth^2
   \right.\right.\right.\right.\nonumber\\
   &+&\left.\left.\left.\left.
5 \sqrt{2} \sth^4 \cth+8 \sth^5\right) M_0^2-\left(11 \sqrt{2} \cth^3+15\sth
   \cth^2+3 \sqrt{2} \sth^2 \cth+\sth^3\right) M_{0\eta
   '}^2\right) M_{0\eta }^2
   \right.\right.\nonumber\\
   &+&\left.\left.
2 \sth^3 \left(11 \sqrt{2} \cth^3+15\sth \cth^2+3 \sqrt{2} \sth^2 \cth+\sth^3\right) M_0^2 M_{0\eta
   '}^6\right] 
  \right.\nonumber\\
   &+&\left.
3 M_{0\pi }^6 \left[\cth \left(\cth+2 \sqrt{2}\sth\right) \left(2
   \cth^2 M_0^2-M_{0\eta '}^2\right) M_{0\eta }^6
   \right.\right.\nonumber\\
   &+&\left.\left.
      2 \cth\sth \left(-\sqrt{2} \cth^2+\sth \cth+\sqrt{2} \sth^2\right) M_0^2 M_{0\eta '}^2
   M_{0\eta }^4
   \right.\right.\nonumber\\
   &+&\left.\left.
\sth M_{0\eta '}^4 \left(2 \cth \left(-\sqrt{2} \cth^2+\sth \cth+\sqrt{2} \sth^2\right) M_0^2+\left(2 \sqrt{2} \cth-\sth\right) M_{0\eta '}^2\right) M_{0\eta }^2+2 \sth^3 \left(\sth-2 \sqrt{2}
   \cth\right) M_0^2 M_{0\eta '}^6\right]
   \right\}\nonumber\\
   &-&\frac{128 M_0^4 (L_8^{r})^2}{27 F^2 M_{0\eta }^6 M_{0\eta '}^6}
   \left\{
    16 M_{0K}^8\left[\cth^2 \left(\cth^4-4 \sqrt{2}\sth \cth^3+12 \sth^2 \cth^2-8 \sqrt{2} \sth^3 \cth+4 \sth^4\right) M_{0\eta }^6
    \right.\right.\nonumber\\
   &+&\left.\left.
   \cth \left(2 \cth^2-\sth^2\right)
   \left(\cth^3-6 \sth^2 \cth+4 \sqrt{2} \sth^3\right)
   M_{0\eta '}^2 M_{0\eta }^4
    \right.\right.\nonumber\\
   &+&\left.\left.
  \sth \left(\cth^2-2 \sth^2\right) \left(4 \sqrt{2} \cth^3+6\sth \cth^2-\sth^3\right) M_{0\eta '}^4 M_{0\eta
   }^2
    \right.\right.\nonumber\\
   &+&\left.\left.
   \sth^2 \left(4 \cth^4+8 \sqrt{2}\sth \cth^3+12 \sth^2 \cth^2+4 \sqrt{2} \sth^3 \cth+\sth^4\right) M_{0\eta '}^6\right]
    \right.\nonumber\\
   &-&\left.
   32 M_{0\pi }^2 M_{0K}^6 \left[\cth^2 \left(\cth^4-4 \sqrt{2} \sth \cth^3+12 \sth^2 \cth^2-8 \sqrt{2} \sth^3 \cth+4 \sth^4\right) M_{0\eta }^6
\right.\right.\nonumber\\
&+&\left.\left.
   \cth \left(2 \cth^2-\sth^2\right) \left(\cth^3-6 \sth^2 \cth+4 \sqrt{2} \sth^3\right) M_{0\eta '}^2 M_{0\eta }^4
   \right.\right.\nonumber\\
&+&\left.\left.
\sth \left(\cth^2-2 \sth^2\right) \left(4 \sqrt{2} \cth^3+6\sth \cth^2-\sth^3\right) M_{0\eta '}^4 M_{0\eta }^2
   \right.\right.\nonumber\\
&+&\left.\left.
   \sth^2 \left(4 \cth^4+8 \sqrt{2}\sth \cth^3+12 \sth^2 \cth^2+4
   \sqrt{2} \sth^3 \cth+\sth^4\right) M_{0\eta '}^6\right] 
   \right.\nonumber\\
&+&\left.
8 M_{0\pi }^4 M_{0K}^4\left[\cth^2 \left(5 \cth^4-14 \sqrt{2}\sth \cth^3+33 \sth^2 \cth^2-22 \sqrt{2} \sth^3 \cth+14 \sth^4\right) M_{0\eta }^6
    \right.\right.\nonumber\\
&+&\left.\left.
   \cth \left(4 \cth^5+3 \sqrt{2}\sth
   \cth^4-29 \sth^2 \cth^3+16 \sqrt{2} \sth^3 \cth^2+9 \sth^4 \cth-11 \sqrt{2} \sth^5\right) M_{0\eta '}^2
   M_{0\eta }^4
    \right.\right.\nonumber\\
&+&\left.\left.
  \sth \left(11 \sqrt{2} \cth^5+9\sth \cth^4-16 \sqrt{2} \sth^2 \cth^3-29 \sth^3 \cth^2-3
   \sqrt{2} \sth^4 \cth+4 \sth^5\right) M_{0\eta '}^4 M_{0\eta }^2
    \right.\right.\nonumber\\
&+&\left.\left.
      \sth^2 \left(14 \cth^4+22 \sqrt{2}\sth \cth^3+33 \sth^2\cth^2+14 \sqrt{2} \sth^3 \cth+5 \sth^4\right) M_{0\eta '}^6\right]
   \right.\nonumber\\
&-&\left.
    24 M_{0\pi }^6 M_{0K}^2\left[\cth^2 \left(\cth^2-2
   \sqrt{2}\sth \cth+2 \sth^2\right) M_{0\eta }^6
   \right.\right.\nonumber\\
&+&\left.\left.
   \cth \sth \left(\sqrt{2} \cth^2-\sth \cth-\sqrt{2} \sth^2\right) M_{0\eta '}^2 M_{0\eta }^4
    \right.\right.\nonumber\\
&+&\left.\left.
   \cth\sth \left(\sqrt{2} \cth^2-\sth \cth-\sqrt{2} \sth^2\right) M_{0\eta '}^4 M_{0\eta }^2+\sth^2 \left(2 \cth^2+2 \sqrt{2}\sth \cth+\sth^2\right) M_{0\eta '}^6\right] 
    \right.\nonumber\\
&+&\left.
9 M_{0\pi }^8 \left(\cth^2 M_{0\eta }^6+\sth^2
   M_{0\eta '}^6\right)  \right\}
    \nonumber\\
&+&
\frac{1}{M_{0\eta }^4 M_{0\eta '}^4}\left\{\left(2 M_{0K}^2+M_{0\pi }^2\right) F^2 v_2^{(2)}\left[ \left(\cth^2
   M_0^2-M_{0\eta '}^2\right) M_{0\eta }^2+ \sth^2 M_0^2 M_{0\eta '}^2\right]^2
    \right.\nonumber\\
&+&\left.
\frac{8 L_6^r M_0^4 \left(2 M_{0K}^2+M_{0\pi }^2\right)}{9}\left[
  \cth^2 \left(2 \left(\cth^2-2 \sqrt{2}\sth
   \cth+2 \sth^2\right) M_{0K}^2
   \right.\right.\right.\nonumber\\
&+&\left.\left.\left.
   \left(\cth^2+4 \sqrt{2}\sth
   \cth-\sth^2\right) M_{0\pi }^2\right) M_{0\eta }^4
   \right.\right.\nonumber\\
&+&\left.\left.
   4 \cth \sth\left(\sqrt{2} \cth^2-\sth \cth-\sqrt{2} \sth^2\right) \left(M_{0K}^2-M_{0\pi }^2\right) M_{0\eta '}^2 M_{0\eta }^2
   \right.\right.\nonumber\\
&+&\left.\left.   
   \sth^2 \left(2 \left(2 \cth^2+2 \sqrt{2}
  \sth \cth+\sth^2\right) M_{0K}^2+\left(-\cth^2-4 \sqrt{2}
  \sth \cth+\sth^2\right) M_{0\pi }^2\right) M_{0\eta '}^4
\right]
\right.\nonumber\\
&+&\left.
\frac{8 C_{19} M_0^4}{3}
\left[
   \cth^2 \left(8 \left(\cth^2-2 \sqrt{2}\sth \cth+2 \sth^2\right) M_{0K}^6
\right.\right.\right.\nonumber\\
&-&\left.\left.\left.   
   12 \left(\cth^2-2 \sqrt{2}\sth
   \cth+2 \sth^2\right) M_{0\pi }^2 M_{0K}^4
   \right.\right.\right.\nonumber\\
&+&\left.\left.\left.
  6 \left(\cth^2-2 \sqrt{2} \sth \cth+2 \sth^2\right) M_{0\pi }^4 M_{0K}^2+\left(\cth^2+4
   \sqrt{2}\sth \cth-\sth^2\right) M_{0\pi }^6\right) M_{0\eta }^4
   \right.\right.\nonumber\\
&+&\left.\left.
4 \cth\sth \left(\sqrt{2} \cth^2-\sth \cth-\sqrt{2} \sth^2\right) \left(4 M_{0K}^6-6 M_{0\pi }^2 M_{0K}^4+3 M_{0\pi }^4 M_{0K}^2-M_{0\pi }^6\right) M_{0\eta '}^2 M_{0\eta }^2
     \right.\right.\nonumber\\
&+&\left.\left.
   \sth^2 \left(8 \left(2 \cth^2+2 \sqrt{2}\sth \cth
  + \sth^2\right) M_{0K}^6-12 \left(2 \cth^2+2 \sqrt{2}\sth \cth+\sth^2\right) M_{0\pi }^2 M_{0K}^4
    \right.\right.\right.\nonumber\\
&+&\left.\left.\left.
   6 \left(2 \cth^2+2 \sqrt{2}\sth \cth+\sth^2\right) M_{0\pi }^4 M_{0K}^2+\left(-\cth^2-4 \sqrt{2} \sth \cth+\sth^2\right) M_{0\pi }^6\right) M_{0\eta '}^4
\right]
 \right.\nonumber\\
&+&\left.
\frac{16 C_{31} M_0^4}{9}\left[
  \cth^2 \left(8 \left(\cth^2-2 \sqrt{2}\sth \cth+2 \sth^2\right) M_{0K}^6
 \right.\right.\right.\nonumber\\
&-&\left.\left.\left.  
   12 \left(\cth^2-2 \sqrt{2}\sth \cth+2 \sth^2\right) M_{0\pi }^2 M_{0K}^4
    \right.\right.\right.\nonumber\\
&+&\left.\left.\left.
   6 \left(\cth^2-2 \sqrt{2} \sth \cth+2 \sth^2\right) M_{0\pi }^4 M_{0K}^2+\left(\cth^2+4 \sqrt{2} \sth \cth-\sth^2\right) M_{0\pi }^6\right) M_{0\eta }^4
    \right.\right.\nonumber\\
&+&\left.\left.
   4 \cth
  \sth \left(\sqrt{2} \cth^2-\sth \cth-\sqrt{2} \sth^2\right) \left(4 M_{0K}^6-6 M_{0\pi }^2 M_{0K}^4+3 M_{0\pi }^4 M_{0K}^2-M_{0\pi }^6\right) M_{0\eta '}^2 M_{0\eta }^2
   \right.\right.\nonumber\\
&+&\left.\left.
   \sth^2
   \left(8 \left(2 \cth^2+2 \sqrt{2}\sth \cth+\sth^2\right)
   M_{0K}^6-12 \left(2 \cth^2+2 \sqrt{2}\sth \cth+\sth^2\right) M_{0\pi }^2 M_{0K}^4
   \right.\right.\right.\nonumber\\
&+&\left.\left.\left.
   6 \left(2 \cth^2+2 \sqrt{2}\sth \cth+\sth^2\right) M_{0\pi }^4 M_{0K}^2+\left(-\cth^2-4 \sqrt{2}\sth \cth+\sth^2\right) M_{0\pi }^6\right) M_{0\eta '}^4\right]
  \right.\nonumber\\
&+&\left.
\frac{8 L_7^r M_0^4}{9}
\left[
   \cth^2 \left(8 \sth^2 \left(M_{0K}^2-M_{0\pi }^2\right){}^2+\cth^2 \left(2 M_{0K}^2+M_{0\pi
   }^2\right){}^2
   \right.\right.\right.\nonumber\\
&+&\left.\left.\left.
   4 \sqrt{2} \cth\sth \left(-2 M_{0K}^4+M_{0\pi }^2 M_{0K}^2+M_{0\pi }^4\right)\right) M_{0\eta }^4
   \right.\right.\nonumber\\
&+&\left.\left.
   2 \cth\sth \left(4 \left(\sqrt{2} \cth^2-\sth \cth-\sqrt{2} \sth^2\right) M_{0K}^4
    \right.\right.\right.\nonumber\\
&+&\left.\left.\left.
   2 \left(-\sqrt{2} \cth^2+10 \sth \cth+\sqrt{2} \sth^2\right) M_{0\pi }^2 M_{0K}^2
    \right.\right.\right.\nonumber\\
&+&\left.\left.\left.
   \left(-2 \sqrt{2} \cth^2-7
  \sth \cth+2 \sqrt{2} \sth^2\right) M_{0\pi }^4\right) M_{0\eta '}^2 M_{0\eta }^2
    \right.\right.\nonumber\\
&+&\left.\left.
   \sth^2 \left(8 \cth^2 \left(M_{0K}^2-M_{0\pi }^2\right){}^2+\sth^2 \left(2 M_{0K}^2+M_{0\pi
   }^2\right){}^2+4 \sqrt{2} \cth\sth \left(2 M_{0K}^4-M_{0\pi }^2 M_{0K}^2-M_{0\pi }^4\right)\right) M_{0\eta
   '}^4\right]
   \right.\nonumber\\
&-&\left.
\frac{8 L_{25}^r M_0^2}{3}\left[
   \cth \left(4 \left(\cth-\sqrt{2} \sth\right) M_{0K}^4-4 \left(\cth-\sqrt{2}\sth\right) M_{0\pi }^2 M_{0K}^2+3 \cth M_{0\pi }^4\right) M_{0\eta }^2
   \right.\right.\nonumber\\
&+&\left.\left.
  \sth \left(4 \left(\sqrt{2} \cth+\sth\right) M_{0K}^4
   \right.\right.\right.\nonumber\\
&+&\left.\left.\left.
   -4 \left(\sqrt{2} \cth+\sth\right) M_{0\pi }^2 M_{0K}^2+3 \sth M_{0\pi }^4\right) M_{0\eta '}^2\right) \left(\left(\cth^2 M_0^2-M_{0\eta '}^2\right) M_{0\eta }^2
  + \sth^2 M_0^2 M_{0\eta '}^2\right]
   \right\}
   \nonumber\\
   &-&\frac{M_0^4}{54 M_{0\eta }^4}
   \left\{
      4 \sth^2 \left(2 \left(2 \mu
   _{\eta } (T)+\mu _{\eta '}(T)\right) \cth^4
    +  2 \sqrt{2}\sth \left(4 \mu _{\eta }(T)-\mu _{\eta '}(T)\right) \cth^3
   \right.\right.\nonumber\\
   &+&\left.\left.
   3 \sth^2 \left(4 \mu _{\eta }(T)-\mu _{\eta '}(T)\right) \cth^2
   +2 \sqrt{2} \sth \left(\left(2 \mu _{\eta }(T)+\mu _{\eta '}(T)\right) \sth^2+3 \mu _K(T)\right) \cth
 \right.\right.\nonumber\\
   &+&\left.\left.
   6 \mu _K(T)
   \sth^2 +\sth^4 \left(\mu _{\eta }(T)+2 \mu _{\eta '}(T)\right)\right) M_{0K}^2
\right.\nonumber\\
   &+&\left.
   \sth^2M_{0\pi }^2 \left(-\left(7 \mu _{\eta }(T)+2 \mu _{\eta '}(T)\right) \cth^4+2 \sqrt{2}\sth \left(\mu
   _{\eta '}(T)-10 \mu _{\eta }(T)\right) \cth^3
\right.\right.\nonumber\\
   &+&\left.\left.   
   3 \left(\left(\mu _{\eta '}(T)-4 \mu _{\eta }(T)\right) \sth^2-2 \mu
   _K(T)+3 \mu _{\pi }(T)\right) \cth^2
\right.\right.\nonumber\\
   &-&\left.\left.      
   2 \sqrt{2}\sth \left(\left(8 \mu _{\eta }(T)+\mu _{\eta '}(T)\right) \sth^2+6 \mu _K(T)+9 \mu _{\pi }(T)\right) \cth+18  \sth^2 \mu _{\pi }(T)
 \right.\right.\nonumber\\
   &+&\left.\left.        
   2 \sth^4 \left(\mu _{\eta }(T)-\mu _{\eta '}(T)\right)\right)
   \right\}
   \nonumber\\
   &-&\frac{M_0^4}{54 M_{0\eta '}^4}
   \left\{
    4 M_{0K}^2\cth^2
   \left[\left(2 \mu _{\eta }(T)+\mu _{\eta '}(T)\right) \cth^4-2 \sqrt{2}\sth \left(\mu _{\eta }(T)+2 \mu _{\eta
   '}(T)\right) \cth^3
   \right.\right.\nonumber\\
   &+&\left.\left.    
   \left(6 \mu _K(T)-3 \sth^2 \left(\mu _{\eta }(T)-4 \mu _{\eta '}(T)\right)\right) \cth^2+2 \sqrt{2}\sth \left(\sth^2 \left(\mu _{\eta }(T)-4 \mu _{\eta '}(T)\right)-3 \mu
   _K(T)\right) \cth
   \right.\right.\nonumber\\
   &+&\left.\left.    
   2 \sth^4 \left(\mu _{\eta }(T)+2 \mu _{\eta '}(T)\right)\right] +M_{0\pi }^2\cth^2\left[2 \left(\mu _{\eta '}(T)-\mu _{\eta }(T)\right) \cth^4
   \right.\right.\nonumber\\
   &+&\left.\left.    
 2 \sqrt{2}\sth \left(\mu
   _{\eta }(T)+8 \mu _{\eta '}(T)\right) \cth^3+3 \left(\left(\mu _{\eta }(T)-4 \mu _{\eta '}(T)\right) \sth^2+6 \mu
   _{\pi }(T)\right) \cth^2
   \right.\right.\nonumber\\
   &+&\left.\left.    
   2 \sqrt{2}\sth \cth\left(-\left(\mu _{\eta }(T)-10 \mu _{\eta '}(T)\right) \sth^2+6 \mu _K(T)+9 \mu _{\pi }(T)\right)
   \right.\right.\nonumber\\
   &-&\left.\left.    
   \sth^2 \left(\left(2 \mu _{\eta }(T)+7 \mu
   _{\eta '}(T)\right) \sth^2+6 \mu _K(T)-9 \mu _{\pi }(T)\right)\right]
   \right\}
   \nonumber\\
  &-& \frac{M_0^4}{54 M_{0\eta }^2 M_{0\eta '}^2}
  \left\{
     8 \cth\sth M_{0K}^2\left[\sqrt{2} \left(2 \mu _{\eta }(T)+\mu _{\eta '}(T)\right) \cth^4+\sth \left(2 \mu _{\eta }(T)-5 \mu _{\eta '}(T)\right) \cth^3
   \right.\right.\nonumber\\
  &+&\left.\left.
   3 \sqrt{2} \left(\left(\mu _{\eta '}(T)-\mu _{\eta }(T)\right) \sth^2+\mu _K(T)\right) \cth^2+\sth \cth\left(\left(2 \mu _{\eta '}(T)-5 \mu _{\eta }(T)\right)
   \sth^2+6 \mu _K\right) 
   \right.\right.\nonumber\\
  &-&\left.\left.
   \sqrt{2} \sth^2 \left(\left(\mu _{\eta }(T)+2 \mu
   _{\eta '}(T)\right) \sth^2+3 \mu _K(T)\right)\right] 
    \right.\nonumber\\
  &+&\left.
   2 \cth\sth M_{0\pi }^2
   \left[\sqrt{2} \left(4 \mu _{\eta }(T)+5 \mu _{\eta '}(T)\right) \sth^4+\cth \left(8 \mu _{\eta }(T)+\mu _{\eta
   '}(T)\right) \sth^3
   \right.\right.\nonumber\\
  &+&\left.\left.
   3 \sqrt{2} \left(\left(\mu _{\eta }(T)-\mu _{\eta '}(T)\right) \cth^2+2 \mu _K(T)+3 \mu _{\pi
   }(T)\right) \sth^2
   \right.\right.\nonumber\\
  &+&\left.\left.
 \sth \cth \left(\left(\mu _{\eta }(T)+8 \mu _{\eta '}(T)\right) \cth^2+6 
   \mu _K(T)+9 \mu _{\pi }(T)\right) 
    \right.\right.\nonumber\\
  &-&\left.\left.
   \sqrt{2} \cth^2 \left(\left(5 \mu _{\eta }(T)+4 \mu _{\eta '}(T)\right) \cth^2+6 \mu _K(T)+9 \mu _{\pi }(T)\right)\right]
  \right\}
  \label{chitopu3nnlo}
 \end{eqnarray}

\end{document}